\documentclass[aps,11pt,prd,notitlepage,tightenlines,nofootinbib,showpacs,superscriptaddress]{revtex4-1}
\usepackage{amsmath, amssymb, amsthm}
\usepackage{mathtools}
\usepackage{mathrsfs}
\usepackage{bm}
\usepackage{slashed}   
\DeclareMathAlphabet{\mathpzc}{OT1}{pzc}{m}{it}

\usepackage{graphicx}
\usepackage{multirow}
\usepackage{tikz}
\usepackage[caption=false]{subfig}
\usepackage{relsize}	
\usepackage{array}
\usepackage{float}
\usepackage{color}
\usepackage{xcolor}
\usepackage{soul}
\usepackage{ulem}
\usepackage{cancel}
\usepackage{hyperref}
\hypersetup{colorlinks=true,linkcolor=blue,anchorcolor=blue,citecolor=red, filecolor=blue,urlcolor=blue,bookmarksnumbered=true,
pdfview=FitB
}

\def\dd{{\mathrm{d}}}

\mathchardef\-="2D

\newcommand{\cev}[1]{\reflectbox{\ensuremath{\vec{\reflectbox{\ensuremath{#1}}}}}}

\colorlet{darkgreen}{green!60!black}
\colorlet{brightyellow}{yellow!75!red}
\colorlet{orange}{red!50!yellow}
\colorlet{darkblue}{blue!60!black}
\colorlet{darkred}{red!80!black}
\colorlet{greenblue}{green!50!blue}

\makeatletter

\newcommand{\Rmnum}[1]{\expandafter\@slowromancap\romannumeral #1@}
\makeatother

\begin{document}
\title{Forces inside a strongly-coupled scalar nucleon}

\author{Xianghui Cao}
\affiliation{Department of Modern Physics, University of Science \& Technology of China, Hefei 230026, China}

\author{Yang Li}
\affiliation{Department of Modern Physics, University of Science \& Technology of China, Hefei 230026, China}

\author{James P. Vary}
\affiliation{Department of Physics and Astronomy, Iowa State University, Ames, IA 50011, USA}
\date{\today}
 \begin{abstract}

We investigate the gravitational form factors of a strongly coupled scalar theory that mimic the interaction between the nucleon and the pion. The non-perturbative calculation is based on the light-front Hamiltonian formalism. We renormalize the energy-momentum tensor with a Fock sector dependent scheme. We also systematically analyze the Lorentz structure of the energy-momentum tensor and identify the suitable hadron matrix elements to extract the form factors, avoiding the contamination of spurious contributions. We verify that the extracted form factors obey momentum conservation as well as the mechanical stability condition. From the gravitational form factors, we compute the energy and pressure distributions of the system. Furthermore, we show that utilizing the Hamiltonian eigenvalue equation, the off-diagonal Fock sector contributions from the interaction term can be converted to diagonal Fock sector contributions, yielding a systematic non-perturbative light-front wave function representation of the energies and forces inside the system. 

 \end{abstract}

\maketitle

\section{Introduction}

One of the biggest mysteries in physics is the nature of the strong force that holds the quarks together inside the nucleons.
This force is responsible for quark confinement, gluon binding, and 99\% of the nucleon mass. The hadronic energy-momentum tensor (EMT) is a direct probe of how the force is distributed inside the hadrons, and has sparked renewed interest in recent research. For recent reviews, see Refs.~\cite{Polyakov:2018zvc, Burkert:2023wzr}. 

By virtue of the Lorentz symmetry, the hadron matrix element of the EMT of the nucleon (spin-1/2) can be parametrized by the gravitational form factors (GFFs) \cite{Kobzarev:1962wt, Pagels:1966zza}, 
\begin{multline}
\langle p', s'|T^{\mu\nu}(0)|p, s\rangle = 
\frac{1}{2M}\overline u_{s'}(p') \Big[
2P^\mu P^{\nu} A(q^2)  
+ iP^{\{\mu}\sigma^{\nu\}\rho}q_\rho J(q^2) \\
+ \frac{1}{2} \big(q^\mu q^\nu-q^2 g^{\mu\nu}\big) D(q^2)
\Big] u_s(p)
\end{multline}
where, $P = (p'+p)/2$, $q = p'-p$, $a^{\{\mu}b^{\nu\}} \equiv  a^\mu b^\nu + a^\nu b^\mu$, and $M^2 = p^2 = p'^2$ is the hadron mass.  
The Lorentz scalars $A(q^2)$, $J(q^2)$ and $D(q^2)$ encode the information about the energy, momentum and stress distributions inside the hadrons. Similarly, for scalar hadrons (spin-0), the hadron matrix elements of the EMT can be written as~\cite{Polyakov:2018zvc}, 
\begin{equation}
\langle p'|T^{\mu\nu}(0)|p\rangle = 
2P^\mu P^{\nu} A(q^2) + \frac{1}{2} \big(q^\mu q^\nu-q^2 g^{\mu\nu}\big) D(q^2).
\end{equation}
The GFFs $A$ and $D$ exist for particles of all spins. 

The $D$ term $D(q^2)$ is associated with the stress component of the EMT, which reveals the mechanical properties of the nucleons. However, this form factor remains poorly understood.
In fact, the $D$ term is dubbed as ``the last global unknown'' \cite{Polyakov:2018zvc}. Unlike other GFFs, the value of the $D$ term at zero-momentum transfer $D(0)$ is not fixed by global conservation laws. It depends on the internal QCD dynamics. Based on mechanical stability, Polyakov et al. conjectured that $D(0)$ should be negative \cite{Perevalova:2016dln}. This is supported by various model calculations \cite{Polyakov:2018zvc, Ji:1997gm, Petrov:1998kf, Polyakov:1998ze, Polyakov:2002yz, Guzey:2005ba, Cebulla:2007ei, Pasquini:2007xz, Hwang:2007tb, Mai:2012yc, Mai:2012cx, Jung:2013bya, Cantara:2015sna, Hudson:2017oul, Polyakov:2018exb, Chakrabarti:2020kdc, Metz:2021lqv, Mamo:2022eui, More:2021stk, Fujita:2022jus, GarciaMartin-Caro:2023klo, More:2023pcy, Freese:2022jlu, Amor-Quiroz:2023rke, Chakrabarti:2023djs}. However, Ji and Liu argued that $D(0)$ is positive for the hydrogen atom, a stable system \cite{Ji:2021mfb}. The $D$ term also contributes to the anomalous mass of the proton, which probes the gluon distribution inside the nucleons \cite{Ji:1994av} (cf.~\cite{Lorce:2018egm, Lorce:2021xku, Burkert:2023wzr}). The $D$ term of the vacuum state gives the cosmological constant \cite{Teryaev:2013qba}. 

In the parton picture, the $D$ term contributes to the generalized parton distributions (GPDs) in the off-forward region \cite{Polyakov:2002yz, Ji:1996nm}, which provide the experimental access to the $D$ term. Burkert et al. extracted the force distributions inside the proton from the deeply virtual Compton scattering data collected at Jefferson lab \cite{Burkert:2018bqq, Burkert:2021ith}. However, the result has large systematic uncertainty and model dependence, due to the limited energy and luminosity of the experiment \cite{Kumericki:2019ddg, Dutrieux:2021nlz}. These measurements are expected to be significantly improved with the advent of electron ion colliders (EICs) \cite{Accardi:2012qut, Anderle:2021wcy, Accardi:2023chb}.  In the time-like region, the $D$ term can also be extracted from the two-photon production of the particle-antiparticle pair, as shown by Kumano et al. \cite{Kumano:2017lhr}. Kharzeev proposed to use the near-threshold quarkonium photo-production to extract the anomalous mass \cite{Kharzeev:2021qkd}. The nucleon $D$ term is also extracted from Lattice QCD \cite{Shanahan:2018nnv, Hackett:2023nkr} and a light-cone sum rule approach \cite{Anikin:2019kwi}. 

The $D$ term has been extensively investigated in perturbation theory, e.g. QCD at large momentum transfer \cite{Tong:2022zax}. However, calculations with non-perturbative couplings are scarce. 
The force distribution has also been explored in phenomenological models \cite{Petrov:1998kf, Polyakov:1998ze, Polyakov:2002yz, Guzey:2005ba, Cebulla:2007ei, Pasquini:2007xz, Hwang:2007tb, Mai:2012yc, Mai:2012cx, Jung:2013bya, Cantara:2015sna, Hudson:2017oul, Polyakov:2018exb, Chakrabarti:2020kdc, Metz:2021lqv, Mamo:2022eui, More:2021stk, Fujita:2022jus, GarciaMartin-Caro:2023klo, More:2023pcy, Freese:2022jlu, Amor-Quiroz:2023rke, Chakrabarti:2023djs}, such as holography \cite{Mamo:2022eui, Fujita:2022jus}, light-front quark-diquark model \cite{Chakrabarti:2020kdc}, and Dyson-Schwinger equations with contact interaction \cite{Xing:2022mvk}. %
One of the challenges of computing the $D$ term is that the stress component of the EMT involves the interaction, which needs to be properly renormalized and consistent with the dynamical solution of the hadron state. Otherwise, the extracted $D$ term may have unphysical divergences, as reported in prior work \cite{Chakrabarti:2020kdc}. 

Light-front quantization is a natural framework to investigate the structures of the hadrons in the non-perturbative regime \cite{Brodsky:1997de, Bakker:2013cea, Gross:2022hyw}. 
It is the native language for parton physics, which describes how the hadrons are seen in modern high energy scattering experiments \cite{Feynman:1969ej, Kogut:1972di, Brodsky:1981rp}. The spatial distribution of hadrons is only meaningful on the light front \cite{Miller:2010nz}. This approach is closely related to physics in the infinite momentum frame \cite{Ji:2021znw}. Furthermore, it also offers a systematic non-perturbative framework to tackle the strong interaction based on the Hamiltonian formalism \cite{Brodsky:1997de}. Remarkable progress has been made in recent years with the development of theoretical methods and the increase of the computational power \cite{Gross:2022hyw, Li:2015zah, Xu:2022abw, Kreshchuk:2020aiq, Qian:2021jxp, Vary:2016ccz}. 
One of the core quantities in light-front quantization is the light-front wave function (LFWF), which encodes the full quantum information of the system \cite{Brodsky:1997de}. 
The LFWF  
representation has been crucial in the investigation of the GFFs $A(q^2)$ and $J(q^2)$. For instance, this representation provides a non-perturbative derivation of the absence of the anomalous gravitomagnetic moment, i.e. $B(0) = 0$ \cite{Brodsky:2000ii}, where $B(q^2) = 2J(q^2) - A(q^2)$. 

In non-relativistic quantum many-body theory, the quantum stress can be expressed as the virial $\sum_i \vec r_i \cdot \vec p_i$ using quantum many-body wave functions \cite{Godfrey:1988, Rogers:2001}. It is natural to expect that such a formulation can be generalized to quantum field theories (QFTs) using LFWFs. One of the main challenges in QFT is the involvement of the interaction terms in the EMT, which change the particle numbers and generate non-diagonal contributions within the Fock space.It has been posited that a LFWF representation of the $D$ term is not useful unless the full set of LFWFs was obtained \cite{Polyakov:2018zvc}. Indeed, in phenomenological models, only the lowest Fock sectors\footnote{Quite often, only the valence sector is retained. See Refs.~\cite{Lan:2021wok, Xu:2022abw, Pasquini:2023aaf} for recent progress in incorporating higher Fock sector contributions.} are available, and the effective quark-antiquark interactions are used. The interaction part of the EMT is effectively neglected. 

In this work, we investigate the GFFs $A(q^2)$ and $D(q^2)$ in a strong-coupling scalar theory. The theory is one of the simplest yet nontrivial QFTs. It can be used to model the low-energy interaction of the (mock) nucleon and the (mock) pion.  This theory is used to illustrate the properties of the $D$ term and its relation to GPDs based leading-order perturbative calculation \cite{Pobylitsa:2002vw, Polyakov:1999gs}. It is interesting to see how the forces inside the mock nucleon evolve as the coupling increases. 
This theory can be systematically renormalized using Fock sector dependent renormalization (FSDR) developed by Karmanov et al. \cite{Karmanov:2008br}, which ensures the cancellation of the UV divergences even in the strong coupling regime. 
The eigenstate in the one-nucleon system is obtained in the light front Hamiltonian approach up to 4 particles (one mock nucleon plus three mock pions), where the Fock sector convergence is achieved \cite{Karmanov:2016yzu, Li:2015iaw}. This theory serves as an ideal playground for exploring the properties of the EMT in the non-perturbative regime. 
We evaluate the hadron matrix elements of the EMT and extract the GFFs. The EMT is renormalized with FSDR. 
From these results, we also obtain a general LFWF representation of the $D$ term, independent of the interaction details. 

The rest of this work is organized as follows. In Sect.~\ref{sec:formalism}, we introduce the light-front Hamiltonian formalism and set up the basic framework for the calculation. In Sect.~\ref{sec:HMEs}, we compute the hadron matrix elements of the EMT. In Sect.~\ref{sec:GFFs}, we first analyze the covariant structure of the hadron matrix elements using covariant light-front dynamics (CLFD) \cite{Carbonell:1998rj}. Based on these analyses and the expressions obtained in Sect.~\ref{sec:HMEs}, we extract the GFFs $A(q^2)$ and $D(q^2)$ and present their numerical results. In Sect.~\ref{sec:LFWF_representation}, we further analyze the hadron matrix elements using the LFWF representation. In the forward limit ($q = 0$), momentum conservations and the Hamiltonian dynamics lead to well-known constraints on the GFFs.  In the off-forward limit ($q \ne 0$), we derive a general LFWF representation for the $A$ term as well as the $D$ term. Finally, we conclude in Sect.~\ref{sec:conclusion}. Some technical details are given in the Appendix.

\section{Light-front Hamiltonian formalism}\label{sec:formalism}

The Lagrangian of the theory reads,
\begin{equation}
    \mathscr L = \partial_\mu \chi^\dagger \partial^\mu \chi - m_0^2 \chi^\dagger\chi 
    + \frac{1}{2} \partial_\mu \varphi \partial^\mu \varphi - \frac{1}{2}\mu^2_0 \varphi^2 + g_0 \chi^\dagger\chi \varphi. 
\end{equation}
Here, $m_0$ and $\mu_0$ are the bare masses of the complex scalar field $\chi$ (mock nucleon) and real scalar field $\varphi$ (mock pion), respectively. The physical masses are taken to be the nucleon mass $m = 0.94\,\mathrm{GeV}$ and the pion mass $\mu = 0.14\,\mathrm{GeV}$, respectively. $g_0$ is the bare coupling. We also introduce a dimensionless coupling, $\alpha = g^2/(16\pi m^2)$, where $g$ is the physical coupling, $m$ is the physical mass of the mock nucleon. In the semi-classical limit, this theory describes a Yukawa type interaction with the coupling strength $\alpha$. 

To simplify our calculations, we quench the theory, i.e., we neglect the contributions of nucleon-antinucleon loops, which would otherwise destabilize the vacuum \cite{Gross:2001ha}. This approximation is justified by the large mass gap between the nucleons and the pions, which suppresses the pair creation effects. 
It is useful to introduce the mass counterterm: $\delta m^2 = m^2 - m_0^2$. In the quenched theory, the mock pion mass is not renormalized, viz. $\mu_0 = \mu$. The Fock space is built with Fock particles with the physical masses. N.B. we do not renormalize the fields, but instead normalize the state vector. The bare parameters are determined in FSDR \cite{Karmanov:2016yzu, Li:2015iaw}. 

The EMT can be obtained as a Noether current of the translational symmetry\footnote{Note that for this simple scalar theory, the EMT is symmetric.},
\begin{equation}\label{eqn:EMT}
    {T}^{\mu\nu} =  \partial^{\{\mu}\chi^\dagger \partial^{\nu\}}\chi  
    - g^{\mu\nu}\big(\partial_\sigma \chi^\dagger\partial^\sigma \chi - m^2_0 \chi^\dagger\chi \big)
    - g^{\mu\nu} g_0 \chi^\dagger \chi\varphi  \\
    + \partial^\mu\varphi \partial^\nu\varphi - \frac{1}{2}g^{\mu\nu} \big(\partial^\rho\varphi\partial_\rho\varphi - \mu^2\varphi^2\big)
\end{equation}
Note that the EMT contains the bare parameters and needs to be renormalized. We adopt the same sector dependent bare parameters determined in FSDR. 
We quantize the theory on the light front $x^0 + x^3 = 0$, and the light-front Hamiltonian is obtained as the conserved charge of $T^{+-}$ \cite{Dirac:1949cp}, 
\begin{equation} \label{eqn:LF_Hamiltonian}
   \mathcal P^- = \int \dd^3x  \, T^{+-} = \int \dd^3 x\, 
    \Big\{
    \chi^\dagger\big[(i\nabla_\perp)^2+m^2 \big] \chi
    + 
    \frac{1}{2}\varphi\big[(i\nabla_\perp)^2+\mu^2 \big]\varphi
    - g_0 \chi^\dagger\chi \varphi - \delta m^2 \chi^\dagger \chi 
    \Big\}
\end{equation}
Here, the light-front component of a 4-vector $v^\mu$ is defined as, 
\begin{equation}
v^\pm = v^0 \pm v^3, \quad \vec v_\perp = (v^1, v^2)
\end{equation}
In particular, $x^+ = x^0 + x^3$ is the light-front time and $p^- = p^0 - p^3$ is the light-front energy. The integration measure 
is defined as $\dd^3 x \equiv \frac{1}{2} \dd x^-\dd^2x_\perp$.

At the initial time $x^+ = 0$, the field operators can be expanded in terms of the creation and annihilation operators, 
\begin{align}
\chi(x) =\,& \int \frac{\dd p^+\dd^2 p_\perp}{(2\pi)^32p^+} 
\Big[b(p) e^{-ip\cdot x} + d^\dagger(p) e^{+ip\cdot x}\Big] \Big|_{x^+=0}, \label{eqn:chi_field} \\
\varphi(x) =\,& \int \frac{\dd p^+\dd^2 p_\perp}{(2\pi)^32p^+} 
\Big[a(p) e^{-ip\cdot x} + a^\dagger(p) e^{+ip\cdot x}\Big] \Big|_{x^+=0}. \label{eqn:phi_field}
\end{align}
The creation and annihilation operators obey the canonical commutation relations,
\begin{equation} \label{eqn:CCR}
\big[a(p), a^\dagger(p')\big] = 2p^+(2\pi)^3\delta^3(p-p'), \quad
\big[b(p), b^\dagger(p')\big] = 2p^+(2\pi)^3\delta^3(p-p').
\end{equation}

The state vectors of the physical particles are the solutions of the light-front Schrödinger equation, 
\begin{equation}\label{eqn:LF_Schrödinger_equation}
    \mathcal P^- |\psi(p)\rangle = \frac{\vec p^2_\perp+M^2}{p^+}|\psi(p)\rangle.
\end{equation}
In this work, we consider the one-nucleon sector, i.e., the mock nucleon dressed by mock pions. The mass eigenvalue is simply the physical nucleon mass $M = m$. 
The state vector can be expressed in the Fock space, 
\begin{equation}\label{eqn:state_vector}
|\psi(p)\rangle = \sum_{n} \int \big[\dd x_i \dd^2 k_{i\perp} \big]_n \psi_n(\{x_i, \vec k_{i\perp}\}) |\{x_i p^+, \vec k_{i\perp}+x_i \vec p_\perp\}_n\rangle,
\end{equation}
where, $[\dd x_i \dd^2 k_{i\perp}]_n = \frac{1}{(n-1)!} \prod_{i=1}^n \frac{\dd x_i}{2x_i}\frac{\dd^2 k_{i\perp}}{(2\pi)^3} 2\delta(\sum_i x_i - 1) (2\pi)^3\delta^2(\sum_i \vec k_{i\perp}) $ is the $n$-body phase space measure. $\psi_n(\{x_i, \vec k_{i\perp}\})$ is called the LFWF, which only depends on the longitudinal momentum fractions $x_i={p_i^+}/{p^+}$ and the relative transverse momenta $\vec k_{i\perp} = \vec p_{i\perp} - x_i \vec p_\perp$.
 And $|\{p_i\}_n\rangle = a^\dagger(p_1)a^\dagger(p_2)\cdots a^\dagger(p_{n-1}) b^\dagger(p_n)|0\rangle$ is the $n$-body Fock state.
Each Fock particle is on their mass shell: $p_i^2 = m_i^2$. The state vector is normalized as $\langle\psi(p')|\psi(p)\rangle=2p^+(2\pi)^3\delta^3(p-p')$. Consequently, the LFWFs are normalized to unity, 
\begin{equation}
\sum_{n}  
\int \big[\dd x_i \dd^2k_{i\perp}\big] \, \psi_n(\{x_i, \vec k_{i\perp}\})\psi_n^*(\{x_i, \vec k_{i\perp}\}) = 1.
\end{equation}

It is useful to introduce the vertex functions $\Gamma_n(\{x_i, \vec k_{i\perp}\}) = (s_n - M^2) \psi_n(\{x_i, \vec k_{i\perp}\})$, where, $s_n = (p_1+p_2+\cdots+p_n)^2 = \sum_i (\vec k_{i\perp}^2+m_i^2)/x_i$ is the invariant mass squared of the Fock sector. It can be shown that the vertex function is the matrix element of the $T$ matrix, 
\begin{equation}\label{eqn:T+-_int}
\langle \{x_i p^+, \vec k_{i\perp}+x_i \vec p_\perp\}_n | T^{+-}_\text{int} (0) |\psi(p)\rangle = -2\Gamma_n(\{x_i, \vec k_{i\perp}\}).
\end{equation}
 Using the vertex functions, we can generalize the light-cone perturbation theory to the non-perturbative regime \cite{Weinberg:1966jm, Karmanov:1976iv, Carbonell:1998rj}. 

The theory is super-renormalizable. However, ultraviolet divergence does exist in loop integrals. We introduce the Pauli-Villars degree of freedom to regularize the divergence prior to renormalizing the theory. After a successful renormalization of the EMT operator, the GFFs are independent of the Pauli-Villars mass, as expected for the conserved current. We will suppress the Pauli-Villars regularization throughout.  
In Ref.~\cite{Li:2015iaw}, the theory is solved in the one-nucleon system up to 4 particles (1 mock nucleon plus 3 mock pions). By comparing with the solution from the three-body truncation \cite{Karmanov:2008br, Karmanov:2016yzu}, the Fock sector expansion is shown to converge up to the three-body truncations for the field strength renormalization constant as well as the electromagnetic form factor up to non-perturbative couplings. In this work, we adopt the three-body truncation to compute the GFFs.

\section{Hadron matrix elements}\label{sec:HMEs}

\begin{figure}
    \centering
    \subfloat[\ $\delta m^2=\delta m_3^2$]{\includegraphics[width=0.23\textwidth]{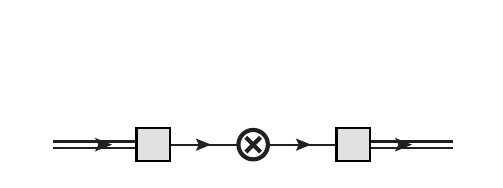}}
    \subfloat[\ $g_0 = g_{03}$]{\includegraphics[width=0.23\textwidth]{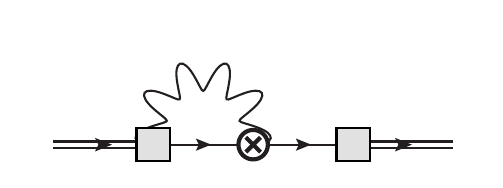}} 
    \subfloat[\ $\delta m^2 = \delta m_2^2$ ]{\includegraphics[width=0.23\textwidth]{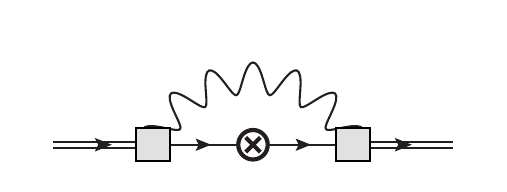}}
    \subfloat[\  ]{\includegraphics[width=0.23\textwidth]{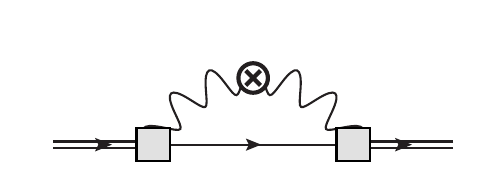}} \\
   \subfloat[\  ]{\includegraphics[width=0.23\textwidth]{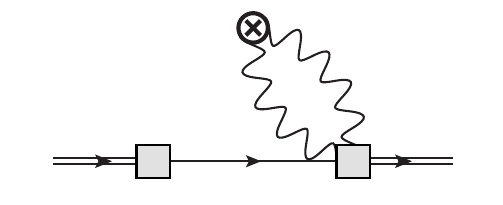}}     \subfloat[\ $g_0 = g_{02}$]{\includegraphics[width=0.23\textwidth]{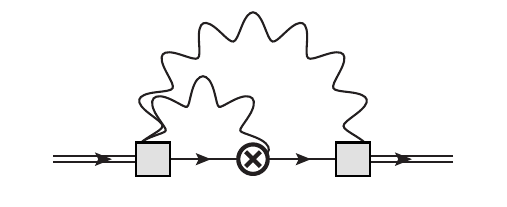}}   
   \subfloat[\ $\delta m=0, g_0 = g_{02}$]{\includegraphics[width=0.23\textwidth]{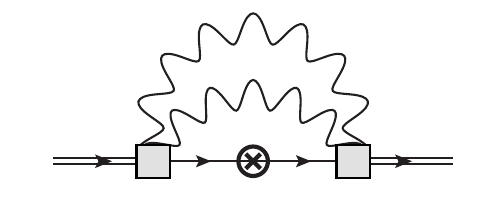}}
    \subfloat[\ $g_0 = g_{02}$]{\includegraphics[width=0.23\textwidth]{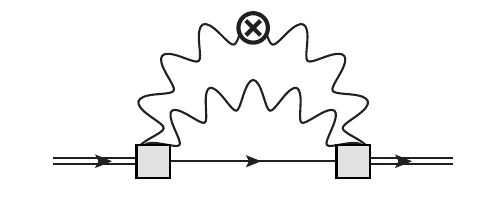}}
    \caption{Diagrammatic representations of the matrix elements of the EMT. The subcaption gives the assignments of the counterterms. Diagrams that are the complex conjugate to these diagrams are not shown. The vertex rules are given in Fig.~\ref{fig:vertex_rules}.}
    \label{fig:emt_ff}
\end{figure}

\begin{figure}
    \centering
    \subfloat[\ $(\frac{1}{2}q^2-\delta m^2)g^{\mu\nu}+p^{\{\mu} p'^{\nu\}} $ ]{\includegraphics[width=0.2\textwidth]{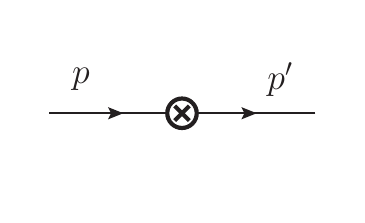}}~
    \subfloat[\ $\frac{1}{2}q^2g^{\mu\nu}+p^{\{\mu} p'^{\nu\}} $ ]{\includegraphics[width=0.2\textwidth]{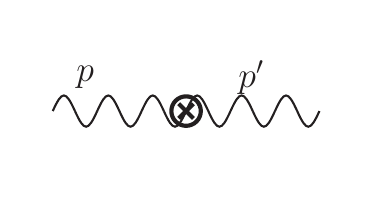}}~
    \subfloat[\  $-g_0g^{\mu\nu}$ ]{\includegraphics[width=0.2\textwidth]{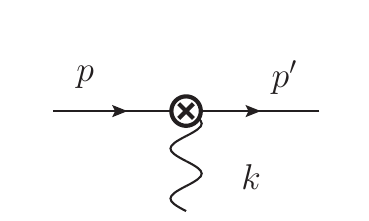}}
    \caption{Vertex rules of the EMT. The light-front time $x^+$ flows from left to right. The solid line represents the nucleon. The wavy lines represent the pion. The circled cross ${\otimes}$ represents the EMT operator. $q$ is the injected momentum.  }
    \label{fig:vertex_rules}
\end{figure}

With the LFWFs solved from the light-front Schrödinger equation (\ref{eqn:LF_Schrödinger_equation}), we can compute the hadron matrix elements of the EMT $t^{\alpha\beta} \equiv \langle p' | T^{\alpha\beta}(0)| p \rangle$ using Eqs.~(\ref{eqn:chi_field}--\ref{eqn:CCR},  \ref{eqn:state_vector} \& \ref{eqn:EMT}). The calculation can be neatly represented using light-front time-ordered diagrams as shown in Fig.~\ref{fig:emt_ff}. See Ref.~\cite{Brodsky:1997de} for a summary of the old-fashioned diagrammatic rules for light-front dynamics, also known as the Weinberg rules \cite{Weinberg:1966jm}. 
The rules for the interaction vertices associated with the EMT can be obtained from the operator expression (\ref{eqn:EMT}), as shown in Fig.~\ref{fig:vertex_rules}. 
These rules are generalized to the non-perturbative regime by adopting the vertex functions for the hadron vertices \cite{Karmanov:1976iv, Carbonell:1998rj}.
It is useful to introduce kinematical variables $P = (p'+p)/2$ and $q = p'-p$. We also adopt the Drell-Yan frame $q^+=0$ to evaluate the hadron matrix elements, which dramatically simplify the algebra \cite{Drell:1969km}. 

\begin{figure}
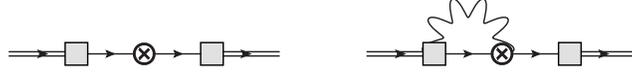

    \centering
    \includegraphics[width=0.275\textwidth]{GFF11a.pdf}~
    \includegraphics[width=0.275\textwidth]{GFF12a.pdf}
    \caption{Diagrams (a) and (b)}
    \label{fig:emt_ab}
\end{figure}

As we mentioned, the EMT contains counterterms. We adopt the FSDR scheme and assign the counterterms according to the Fock sector in which they reside. The values of the counterterms were obtained from solving the Schrödinger equation in the previous work~\cite{Li:2015iaw, Karmanov:2016yzu}. These counterterms are sufficient to renormalize the EMT. To see this point, let us first see an example. Consider diagrams (a) \& (b), as shown in Fig.~\ref{fig:emt_ab}. The corresponding expressions are, 
\begin{align}
t_a^{\alpha\beta} =\,& Z\big[ (\frac{1}{2}q^2-\delta m_3^2)g^{\alpha\beta} + p^{\{\alpha}p'^{\beta\}}\big] \nonumber \\
=\,& Z\big[2P^\alpha P^\beta + (\frac{1}{2}q^2-\delta m_3^2)g^{\alpha\beta} - \frac{1}{2}q^{\alpha}q^\beta\big], \\
t^{\alpha\beta}_b =\,& -\sqrt{Z}g^{\alpha\beta}\int\frac{\dd x}{2x(1-x)}\int \frac{\dd^2k_\perp}{(2\pi)^3} g_{03} \psi_2(x, k_\perp).
\end{align}
Here, $\sqrt{Z} = \psi_1$ is the one-body normalization constant. $t_a$ contains a mass counterterm $\delta m_3^2$. 
Compare $t_b$ with the self-energy function within the 3-body truncation, 
\begin{equation}
\Sigma(m^2) \equiv \Sigma^{(3)}(m^2) = - \frac{1}{\sqrt{Z}}\int \frac{\dd x}{2x(1-x)}\int \frac{\dd^2 k_\perp}{(2\pi)^3} g_{03} \psi_2(x, k_\perp) = \delta m^2_3
\end{equation}
We obtain, 
\begin{equation}
t^{\alpha\beta}_b = g^{\alpha\beta} Z \delta m^2_3 
\end{equation}
as expected. 
Therefore, $t_b$ cancels out the mass counterterm contribution in diagram (a), 
\begin{equation}
t_a^{\alpha\beta} + t^{\alpha\beta}_b  = Z\big[2P^\alpha P^\beta - \frac{1}{2}q^2_\perp g^{\alpha\beta} - \frac{1}{2}q^{\alpha}q^\beta\big]
\end{equation}

We collect the result of each diagram as follows. 

\subsection{Diagram (c)}

\begin{figure}[h]
    \centering
    \includegraphics[width=0.3\textwidth]{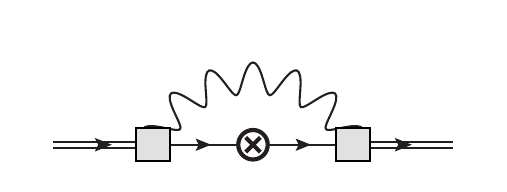}
    \caption{Diagram (c): $\delta m^2 = \delta m_2^2$}
    \label{fig:emt_c}
\end{figure}

The relevant diagram is shown in Fig.~\ref{fig:emt_c}.
\begin{align}
t^{\alpha\beta}_c =\,& \int \frac{\dd x}{2x (1-x)^2}\int \frac{\dd^2k_\perp}{(2\pi)^3} \psi_2(x, \vec k_\perp) \psi_2^*(x, \vec k_\perp-x\vec q_\perp) \big[ (\frac{1}{2}q^2 - \delta m_2^2)g^{\alpha\beta} + p_1^{\{\alpha}p'^{\beta\}}_1\big]
\end{align}
where, $p_1, p'_1$ are, 
\begin{align}
p_1^+ =\,& (1-x)P^+ \\
\vec p_{1\perp} =\,& -\vec k_\perp + (1-x)\vec P_\perp - \frac{1}{2}(1-x)\vec q_\perp \\
p_1^- =\,& \frac{p_{1\perp}^2+m^2}{p_1^+} \\
p'^+_1 =\,& (1-x)P^+ \\
\vec p'_{1\perp} =\,& -\vec k_\perp + (1-x)\vec P_\perp + \frac{1}{2}(1+x)\vec q_\perp \\
p'^-_1 =\,& \frac{p'^2_{1\perp}+m^2}{p'^+_1}
\end{align}

\subsection{Diagram (d)}

\begin{figure}[h]
    \centering
    \includegraphics[width=0.3\textwidth]{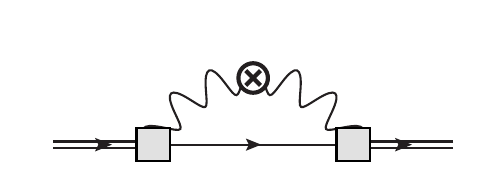} 
    \caption{Diagram (d)}
    \label{fig:emt_d}
\end{figure}

The relevant diagram is shown in Fig.~\ref{fig:emt_d}.
\begin{equation}
t^{\alpha\beta}_d = \int \frac{\dd x}{2x^2(1-x)} \int \frac{\dd^2k_\perp}{(2\pi)^3}\psi_2(x, \vec k_\perp) \psi_2^*(x, \vec k_\perp + (1-x)\vec q_\perp) (\frac{1}{2} q^2 g^{\alpha\beta} + k_1^{\{\alpha} k'^{\beta\}}_1)
\end{equation}
where, $k_1, k'_1$ are, 
\begin{align}
k_1^+ =\,& xP^+ \\
\vec k_{1\perp} =\,& \vec k_\perp + x\vec P_\perp - \frac{1}{2}x\vec q_\perp \\
k_1^- =\,& \frac{k_{1\perp}^2+\mu^2}{k_1^+} \\
k'^+_1 =\,& xP^+ \\
\vec k'_{1\perp} =\,& \vec k_\perp + x\vec P_\perp + (1-\frac{1}{2}x)\vec q_\perp \\
k'^-_1 =\,& \frac{k'^2_{1\perp}+\mu^2}{k'^+_1}
\end{align}

\subsection{Diagram (e)}

\begin{figure}[h]
    \centering
   \includegraphics[width=0.3\textwidth]{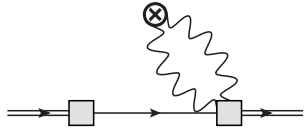}
    \caption{Diagram (e) vanishes in the Drell-Yan frame}
    \label{fig:emt_e}
\end{figure}

The relevant diagram is shown in Fig.~\ref{fig:emt_e}.
In the Drell-Yan frame $q^+=0$, this diagram and its complex conjugate (denoted as $\bar e$) vanish, $t_e =  t_{\bar e} = 0$.

\subsection{Diagram (f)}

\begin{figure}[h]
    \centering
   \includegraphics[width=0.3\textwidth]{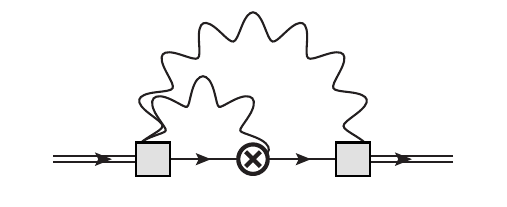}
    \caption{Diagram (f): $g_0 = g_{02}$}
    \label{fig:emt_f}
\end{figure}

The relevant diagram is shown in Fig.~\ref{fig:emt_f}.
\begin{multline}
t^{\alpha\beta}_f = -g_{02}g^{\alpha\beta} \int \frac{\dd x}{2x(1-x)} \int \frac{\dd^2k_\perp}{(2\pi)^3} 
\int \frac{\dd x'}{2x'(1-x-x')} \int \frac{\dd^2k'_\perp}{(2\pi)^3} \\
\times \psi_3(x, \vec k_\perp, x', \vec k'_\perp) \psi_2^*(x, \vec k_\perp - x\vec q_\perp)
\end{multline}

\begin{figure}[h]
    \centering
   \includegraphics[width=0.3\textwidth]{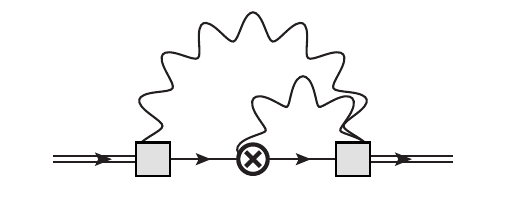}
    \caption{Diagram ($\bar{\text{f}}$): $g_0 = g_{02}$}
    \label{fig:emt_fbar}
\end{figure}

A related diagram, ($\bar{\text{f}}$), is the Hermitian conjugate of diagram (f), as shown in Fig.~\ref{fig:emt_fbar},
\begin{multline}
t^{\alpha\beta}_{\bar f} = -g_{02}g^{\alpha\beta} \int \frac{\dd x}{2x(1-x)} \int \frac{\dd^2k_\perp}{(2\pi)^3} 
\int \frac{\dd x'}{2x'(1-x-x')} \int \frac{\dd^2k'_\perp}{(2\pi)^3} \\
\times \psi_3^*(x, \vec k_\perp-x\vec q_\perp, x', \vec k'_\perp-x'\vec q_\perp) \psi_2(x, \vec k_\perp)
= \bar t_f^{\alpha\beta} 
\end{multline}

\subsection{Diagram (g)}

\begin{figure}[h]
    \centering
   \includegraphics[width=0.3\textwidth]{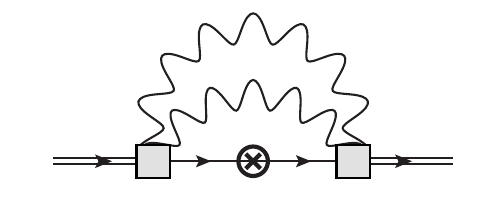}
    \caption{Diagram (g): $\delta m=0, g_0 = g_{02}$}
    \label{fig:emt_g}
\end{figure}

The relevant diagram is shown in Fig.~\ref{fig:emt_g}, 
\begin{multline}
t^{\alpha\beta}_g = \frac{1}{2!} \int \frac{\dd x}{2x}\int \frac{\dd^2 k_\perp}{(2\pi)^3}\int \frac{\dd x'}{2x'(1-x-x')^2}
\int \frac{\dd^2 k'_\perp}{(2\pi)^3} \\
\times \psi_3(x, \vec k_\perp, x', \vec k'_\perp) \psi_3^*(x, \vec k_\perp - x\vec q_\perp, x', \vec k'_\perp - x'\vec q_\perp) \big( \frac{1}{2}q^2 g^{\alpha\beta} + p_1^{\{\alpha} p'^{\beta\}}_1\big) 
\end{multline}
where, 
\begin{align}
p_1^+ =\,& (1-x-x')P^+, \\
\vec p_{1\perp} =\,& -\vec k_\perp - \vec k'_\perp + (1-x-x')\vec P_\perp - \frac{1}{2}(1-x-x')\vec q_\perp, \\
p^-_1 =\,& \frac{ p_{1\perp}^2 + m^2}{p_1^+}; \\
p'^+_1 =\,& (1-x-x')P^+, \\
\vec p'_{1\perp} =\,& -\vec k_\perp - \vec k'_\perp + (1-x-x')\vec P_\perp + \frac{1}{2}(1+x+x')\vec q_\perp, \\
p'^-_1 =\,& \frac{ p'^2_{1\perp} + m^2}{p'^+_1}.
\end{align}

\subsection{Diagram (h)}

\begin{figure}[h]
    \centering
   \includegraphics[width=0.3\textwidth]{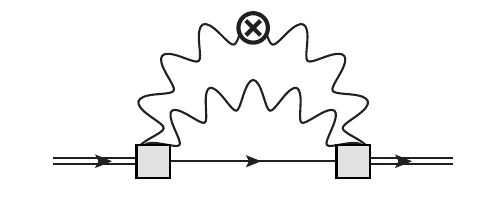}
    \caption{Diagram (h): $\delta m=0, g_0 = g_{02}$}
    \label{fig:emt_h}
\end{figure}

The relevant diagram is shown in Fig.~\ref{fig:emt_h}, 
\begin{multline}
t^{\alpha\beta}_h =  \int \frac{\dd x}{2x}\int \frac{\dd^2 k_\perp}{(2\pi)^3}\int\frac{\dd x'}{2x'^2(1-x-x')} \int \frac{\dd^2k'_\perp}{(2\pi)^3} \\
\times \psi_3(x, \vec k_\perp, x', \vec k'_\perp) \psi_3^*(x, \vec k_\perp + (1- x) \vec q_\perp, x', \vec k'_\perp -x' \vec q_\perp)
\big(\frac{1}{2}q^2g^{\alpha\beta}+k_1^{\{\alpha}k'^{\beta\}}_1\big)
\end{multline}
where, $k_1, k'_1$ are, 
\begin{align}
k^+_1 =\,& xP^+, \\
\vec k_{1\perp} =\,& \vec k_\perp + x\vec P-\frac{1}{2}x\vec q_\perp, \\
k^-_1=\,& \frac{k^2_{1\perp}+\mu^2}{k^+_1}, \\
k'^+_1 =\,& xP^+, \\
\vec k'_{1\perp} =\,& \vec k_\perp + x \vec P + (1-\frac{1}{2}x)\vec q_\perp, \\
k^-_1=\,& \frac{k'^2_{1\perp}+\mu^2}{k'^+_1}. 
\end{align}

\subsection{Forward limit}\label{sec:forward_limit}

To see the renormalization of the EMT, let us first examine the hadron matrix elements in the forward limit ($q = 0$). As we will see later in Sect.~\ref{sec:GFFs}, to extract the GFFs, it is sufficient to consider $t^{++}$ and $t^{+-}$. 

In the LFWF representation, $t^{++}_a = Z2(P^+)^2 = I_12(P^+)^2$, where $I_1=Z$ is the probability of the one-body Fock sector. 
On the other hand, 
\begin{align}
t^{++}_c =\,& 2(P^+)^2\int \frac{\dd x}{2x (1-x)}\int \frac{\dd^2k_\perp}{(2\pi)^3} \psi_2(x, \vec k_\perp) \psi_2^*(x, \vec k_\perp-x\vec q_\perp) (1-x) \\
t^{++}_d =\,& 2(P^+)^2\int \frac{\dd x}{2x (1-x)}\int \frac{\dd^2k_\perp}{(2\pi)^3} \psi_2(x, \vec k_\perp) \psi_2^*(x, \vec k_\perp+(1-x)\vec q_\perp) x
\end{align}
Hence, in the forward limit,  
\begin{equation}
 t^{++}_c + t^{++}_d = I_2 2(P^+)^2
\end{equation}
where, 
\begin{equation}
I_2 = \int \frac{\dd x}{2x(1-x)} \int \frac{\dd^2k_\perp}{(2\pi)^3}\psi_2(x, \vec k_\perp) \psi_2^*(x, \vec k_\perp),
\end{equation}
is the two-body normalization constant. Similarly, one obtains, $t^{++}_g + t^{++}_h = I_3 2(P^+)^2$, where, 
\begin{equation}
I_3 =  \frac{1}{2!}\int \frac{\dd x}{2x}\int \frac{\dd^2 k_\perp}{(2\pi)^3}\int\frac{\dd x'}{2x'(1-x-x')} \int \frac{\dd^2k'_\perp}{(2\pi)^3}  \psi_3(x, \vec k_\perp, x', \vec k'_\perp) \psi_3^*(x, \vec k_\perp, x', \vec k'_\perp),
\end{equation}
is the three-body normalization constant. Note that $t^{++}_b = t^{++}_e = t^{++}_f = 0$. Therefore, in the forward limit,  
\begin{equation}\label{eqn:forward_t++}
t^{++} = (I_1 + I_2 + I_3) 2(P^+)^2 = 2(P^+)^2.
\end{equation}
Here, we have used the normalization condition for the three-body truncation, $I_1 + I_2 + I_3 = 1$. 

Next, let us consider $t^{+-}$. This hadron matrix element is more complicated since it involves the interaction. 
The one-body part of $t^{+-}$ is:
\begin{equation}
t^{+-}_1 \equiv t^{+-}_a + t^{+-}_b = Z(2P^+P^-) = Z(2M^2 + 2P^2_\perp)
\end{equation}
For the 2-body part, let us first consider the kinematical contribution, viz. excluding terms proportional to $g^{\alpha\beta}$,
\begin{equation}
t_{2,kin}^{+-} \equiv t^{+-}_{c,kin} + t^{+-}_{d,kin} = 
 \int \frac{\dd x}{2x (1-x)}\int \frac{\dd^2k_\perp}{(2\pi)^3} \psi_2(x, \vec k_\perp) \psi_2^*(x, \vec k_\perp) 
  \Big[2 \frac{k_\perp^2+\mu^2}{x} + 2 \frac{k_\perp^2+m^2}{1-x} + 2 P_\perp^2\Big]
\end{equation}
Similarly, the 3-body kinematical contributions:
\begin{multline}
t_{3,kin}^{+-} \equiv t^{+-}_{g,kin} + t^{+-}_{h,kin} = \frac{1}{2!} \int \frac{\dd x}{2x}\int \frac{\dd^2 k_\perp}{(2\pi)^3}\int \frac{\dd x'}{2x'(1-x-x')}
\int \frac{\dd^2 k'_\perp}{(2\pi)^3} 
 \psi_3(x, \vec k_\perp, x', \vec k'_\perp) \psi_3^*(x, \vec k_\perp, x', \vec k'_\perp) \\
\times \Big( 
2 \frac{(\vec k_\perp + \vec k'_\perp)^2+m^2}{1-x-x'} + 2 \frac{k_\perp^2+\mu^2}{x} + 2 \frac{k'^2_\perp+\mu^2}{x'} + 2P^2_\perp 
\Big) 
\end{multline}

\begin{figure}[h]
    \centering
   \includegraphics[width=0.9\textwidth]{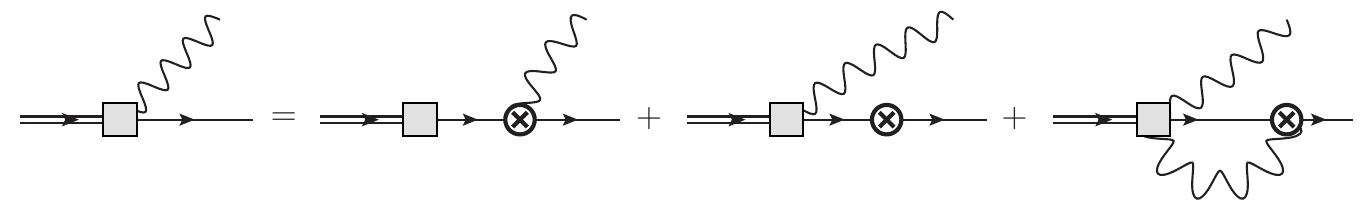}
    \caption{The equation of motion for the two-body vertex function \cite{Karmanov:2016yzu}.}
    \label{fig:emt_eom_2body}
\end{figure}

Next, let us consider the interacting terms, viz. all terms proportional to $g^{\alpha\beta}$, in the forward limit. These involve diagrams
$t_{\bar b}, t_c, t_f, t_{\bar f}, t_g, t_h$. Note that the $g^{\alpha\beta}$ parts of $t_a$ and $t_b$ cancel out. 
The two-body interacting contribution, $t_{\bar b}, t_c, t_f$ reads: 
\begin{multline}\label{eqn:int_2body}
t^{\alpha\beta}_{2,int} \equiv t^{\alpha\beta}_{\bar b} + t^{\alpha\beta}_{c, int} + t^{\alpha\beta}_{f, int}
= - g^{\alpha\beta} \int \frac{\dd x}{2x(1-x)}\int\frac{\dd^2k_\perp}{(2\pi)^3} \psi^*_2(x, \vec k_\perp) \\
\times \bigg\{\sqrt{Z} g_{03} + \frac{1}{1-x}\delta m^2\psi_2(x, \vec k_\perp)
+ g_{02}\int \frac{\dd x'}{2x'(1-x-x')}\int \frac{\dd^2k'_\perp}{(2\pi)^3}
\psi_3(x, \vec k_\perp, x', \vec k'_\perp) 
\bigg\}
\end{multline}
The expression in the curly bracket can be simplified using the equation of motion as shown in Fig.~\ref{fig:emt_eom_2body}:
\begin{equation}
\Gamma_2(x, \vec k_\perp) = \sqrt{Z} g_{03} + \frac{1}{1-x}\delta m^2\psi_2(x, \vec k_\perp)
+ g_{02}\int \frac{\dd x'}{2x'(1-x-x')}\int \frac{\dd^2k'_\perp}{(2\pi)^3}
\psi_3(x, \vec k_\perp, x', \vec k'_\perp) 
\end{equation}
Hence, Eq.~(\ref{eqn:int_2body}) becomes, 
\begin{equation}\label{eqn:reduction}
t^{\alpha\beta}_{2, int} = g^{\alpha\beta} \int \frac{\dd x}{2x(1-x)}\int\frac{\dd^2k_\perp}{(2\pi)^3} \psi^*_2(x, \vec k_\perp) \psi_2(x, \vec k_\perp) 
 \Big(M^2 - \frac{k_\perp^2+\mu^2}{x} - \frac{k_\perp^2+m^2}{1-x}\Big).
\end{equation}
The full 2-body contribution becomes, 
\begin{multline}
t_2^{+-} = t^{+-}_{2, kin} + t^{+-}_{2, int} =  \int \frac{\dd x}{2x (1-x)}\int \frac{\dd^2k_\perp}{(2\pi)^3} \psi_2(x, \vec k_\perp) \psi_2^*(x, \vec k_\perp) 
  \big[2M^2+ 2 P_\perp^2\big] \\
  = I_2 (2m^2+2P^2_\perp)
\end{multline}

\begin{figure}[h]
    \centering
   \includegraphics[width=0.7\textwidth]{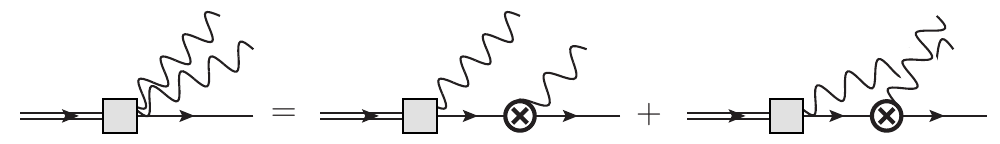}
    \caption{The equation of motion for the three-body vertex function \cite{Karmanov:2016yzu}.}
    \label{fig:emt_eom_3body}
\end{figure}

Finally, the only 3-body contribution of the interacting part is $t^{+-}_{\bar f}$, 
\begin{align}
t^{\alpha\beta}_{\bar f} =\,& - g^{\alpha\beta}\int \frac{\dd x}{2x(1-x)}\int \frac{\dd^2k_\perp}{(2\pi)^3} \int \frac{\dd x'}{2x'(1-x-x')}
\int \frac{\dd^2k'_\perp}{(2\pi)^3} 
\psi^*_3(x, \vec k_\perp, x', \vec k'_\perp)  
  g_{02}\psi_2(x, \vec k_\perp) \\
=\,& - g^{\alpha\beta} \frac{1}{2!} \int \frac{\dd x}{2x}\int \frac{\dd^2k_\perp}{(2\pi)^3} \int \frac{\dd x'}{2x'(1-x-x')}
\int \frac{\dd^2k'_\perp}{(2\pi)^3} 
\psi^*_3(x, \vec k_\perp, x', \vec k'_\perp) \nonumber \\
& \quad \times \Big\{ \frac{g_{02}}{1-x} \psi_2(x, \vec k_\perp) +  \frac{g_{02}}{1-x'} \psi_2(x', \vec k'_\perp) \Big\}.
\end{align}
Again, applying the equation of motion (Fig.~\ref{fig:emt_eom_3body}), the expression in the curly bracket becomes, 
\begin{equation}
\Gamma_3(x, \vec k_\perp, x', \vec k'_\perp ) = 
\frac{g_{02}}{1-x} \psi_2(x, \vec k_\perp) +  \frac{g_{02}}{1-x'} \psi_2(x', \vec k'_\perp).
\end{equation}
The interacting part of diagram-$\bar f$ is, 
\begin{multline}
t^{+-}_{\bar f} = - 2 \frac{1}{2!}\int \frac{\dd x}{2x}\int \frac{\dd^2k_\perp}{(2\pi)^3} \int \frac{\dd x'}{2x'(1-x-x')}
\int \frac{\dd^2k'_\perp}{(2\pi)^3} 
 \psi^*_3(x, \vec k_\perp, x', \vec k'_\perp ) 
\psi_3(x, \vec k_\perp, x', \vec k'_\perp ) \\
\times\Big[ \frac{(\vec k_\perp + \vec k'_\perp)^2+m^2}{1-x-x'} + \frac{k_\perp^2+\mu^2}{x} +  \frac{k'^2_\perp+\mu^2}{x'} -M^2
\Big]
\end{multline}
Therefore, the full 3-body contribution is, 
\begin{equation}
t_3^{+-} = t_{3,kin}^{+-} + t_{\bar f}^{+-} = I_3 (2m^2+2P^2_\perp)
\end{equation}
Summing over all three Fock sector contributions, the full EMT is, 
\begin{equation}\label{eqn:forward_t+-}
t^{+-} =  t^{+-}_1 + t^{+-}_2 + t^{+-}_3  = 2(m^2+P^2_\perp) (I_1 + I_2 + I_3) = 2(m^2+P^2_\perp),
\end{equation}

As one can see, the hadron matrix elements do not depend on any counterterms, nor do they contain any additional divergence. 

\section{Gravitational form factors}\label{sec:GFFs}

\subsection{Covariant light-front dynamics}

By virtue of the Lorentz symmetry,  the hadron matrix elements of the EMT for a scalar particle can be parametrized by 2 GFFs,
\begin{equation}
\langle p' | T^{\alpha\beta}(0)|p\rangle 
= 2 P^\alpha P^\beta A(q^2) + \frac{1}{2}(q^\alpha q^\beta - q^2 g^{\alpha\beta})D(q^2). 
\end{equation}
where, $q = p'-p$ and $P = \frac{1}{2}(p'+p)$. In practical non-perturbative calculations, including those on the light front, however, the full Lorentz symmetry can only be retained in the exact continuum limit. As we impose truncations and approximations (e.g., neglecting higher Fock sectors), the dynamical symmetries, i.e. symmetries involving interactions, are likely to be broken and only the kinematical symmetries are manifest. As a result, the hadron matrix element of the EMT has to be reparametrized with the reduced symmetries \cite{Carbonell:1998rj}. 

In CLFD, this can be systematically constructed by introducing a null vector $\omega^\mu$, which indicates the orientation of the quantization surface, the light front \cite{Carbonell:1998rj}. Effectively, the state vector $|\psi(p)\rangle$ depends on $\omega^\mu$.  The standard light-front coordinate is recovered by choosing $\omega = (1, 0, 0, -1)$, viz. $\omega^- = 2, \omega^+ = \omega_\perp = 0$. Note that since the null vector $\omega$ is only defined up to a scaling factor, the dependence on $\omega$ is always in the form $\omega^\mu / (\omega \cdot P)$. 

In CLFD, the form factors $F_i(\zeta, q^2)$ are complex functions of two Lorentz scalars, $q^2$ and $\zeta = (\omega\cdot q)/(\omega \cdot P)$. Hermiticity of the EMT operator implies $F_i(\zeta, q^2) = F_i^*(-\zeta, q^2)$. In the Drell-Yan frame $\zeta = 0$ \cite{Drell:1969km}, the form factors become real functions. 
In this frame, the most general Lorentz structures of the hadron matrix element read, 
 \begin{multline}
t^{\alpha\beta} = \langle p' | T^{\alpha\beta}(0)|p\rangle 
= 2 P^\alpha P^\beta A(q^2) + \frac{1}{2}(q^\alpha q^\beta - q^2 g^{\alpha\beta})D(q^2) + \frac{(q^2)^2\omega^\alpha\omega^\beta}{(\omega\cdot P)^2}S_1(q^2) \\
+ \frac{1}{(\omega\cdot P)^2} \varepsilon^{\alpha\mu\nu\gamma} P_\mu q_\nu \omega_\gamma \varepsilon^{\beta\rho\sigma\lambda} P_\rho q_\sigma\omega_\lambda S_2(q^2)
\end{multline}
where, $q = p'-p$ and $P = \frac{1}{2}(p'+p)$.  
The new GFFs $S_{1,2}(q^2)$ are the spurious form factors. They are expected to vanish in the continuum limit when the full dynamics is incorporated. 
To extract the GFFs, we can compute the components of the EMT as follows. 
\begin{align}
& t^{++} = 2(P^{+})^2 A(-q^2_\perp) \\
& t^{+i} = 2P^+P^i A(-q^2_\perp) \\
& t^{ij} = 2 P^i P^j A(-q^2_\perp) + \frac{1}{2}(q^i q^j - \delta^{ij}q^2_\perp) D(-q^2_\perp) + (\hat z\times \vec q_\perp)^i (\hat z\times \vec q_\perp)^j S_2(-q_\perp^2) \\
& t^{+-}= 2(m^2+P_\perp^2+\frac{1}{4}q_\perp^2)A(-q^2_\perp) + q^2_\perp D(-q^2_\perp) \label{eqn:t+-} \\
& t^{--} = 8 \Big(\frac{m^2+P^2_\perp+\frac{1}{4}q_\perp^2}{P^+}\Big)^2 A(-q_\perp^2) + 2 \Big(\frac{\vec q_\perp \cdot \vec P_\perp}{P^+}\Big)^2D(-q^2_\perp) + 4 \frac{q^4_\perp}{(P^+)^2}S_1(-q_\perp^2) \\
& \quad + 4\Big(\frac{(\vec P_\perp\times \vec q_\perp)\cdot\hat z}{P^+}\Big)^2S_2(-q_\perp^2) \nonumber \\
& t^{-i} = \frac{m^2+P^2_\perp+\frac{1}{4}q_\perp^2}{P^+}2P^i A(-q_\perp^2) - \frac{\vec q_\perp \cdot \vec P_\perp}{2P^+} q^i D(-q^2_\perp) + \frac{2(\vec P_\perp \times \vec q_\perp)\cdot\hat z}{P^+}\epsilon^{ij} q^j S_2(-q^2_\perp)
\end{align}
Further simplification can be achieved by taking the Breit frame, $\vec P_\perp = \frac{1}{2}(\vec p_\perp + \vec p'_\perp) = 0$,
\begin{align}
& t^{++} = 2(P^{+})^2 A(-q^2_\perp) \\
& t^{+i} = 0 \\
& t^{ij} = \frac{1}{2}(q^i q^j - \delta^{ij}q^2_\perp) D(-q^2_\perp) + \epsilon^{in}\epsilon^{jm} q^n_\perp q^m_\perp S_2(-q_\perp^2) \\
& \mathrm{tr} \tensor t_{\perp\perp} = t^{11} + t^{22} = - \frac{1}{2} q^2_\perp D(-q^2_\perp) + q^2_\perp S_2(-q_\perp^2) \\
& t^{12} = \frac{1}{2} q^1 q^2  D(-q^2_\perp) - q^1 q^2 S_2(-q_\perp^2) \\
& t^{+-}= 2(m^2+\frac{1}{4}q_\perp^2)A(-q^2_\perp) + q^2_\perp D(-q^2_\perp) \\
& t^{--} = 8 \Big(\frac{m^2+\frac{1}{4}q_\perp^2}{P^+}\Big)^2 A(-q_\perp^2) + 4 \frac{q^4_\perp}{(P^+)^2}S_1(-q_\perp^2) \\
& t^{-i} = 0
\end{align}
From these analyses, the $A$ term can be extracted from $t^{++}$, while the $D$ term can be extracted from $t^{+-}$,
\begin{align}
A(-q_\perp^2) =\,& \frac{t^{++}}{2(P^+)^2}, \label{eqn:GFF_A}\\
q_\perp^2 D(-q^2_\perp) =\,& t^{+-}   - \frac{m^2+\frac{1}{4}q_\perp^2}{(P^+)^2} t^{++} \label{eqn:GFF_D}
\end{align}
Note that the popular choice $t^{ij}$, the transverse stress tensor, is contaminated by the spurious GFF $S_2$. Using these components to extract the $D$ term is not reliable, unless a proper combination is used \cite{Chakrabarti:2020kdc}.

\subsection{Physical densities} 

From these two GFFs, we can compute several physical densities. These physical densities attain the same physical interpretation as those in 
classical field theory \cite{Li:2023}, 
\begin{equation}
t^{\alpha\beta} = (e+p) u^\alpha u^\beta - p g^{\alpha\beta} + \pi^{\alpha\beta}.
\end{equation}
Here, $e$ is the proper energy density. $p$ is the pressure, trace part of the stress tensor. Both quantities are Lorentz invariants. $\pi^{\alpha\beta}$ is the (traceless) shear tensor. 
The energy density is defined as \cite{Li:2023},
\begin{equation}
e(r_\perp) = M \int \frac{\dd^2 q_\perp}{(2\pi)^2} e^{-i\vec q_\perp \cdot \vec r_\perp} \Big\{ A(-q_\perp^2) + \frac{q_\perp^2}{4M^2} \big[A(-q_\perp^2) + D(-q_\perp^2)\big] \Big\}
\end{equation}
Note that the energy density is different from the Fourier transform of $t^{+-}$. 
Similarly, the pressure is \cite{Li:2023}, 
\begin{equation}
p(r_\perp) = -\frac{1}{6M} \int \frac{\dd^2 q_\perp}{(2\pi)^2} e^{-i\vec q_\perp \cdot \vec r_\perp} q_\perp^2 D(-q_\perp^2). 
\end{equation}
The shear tensor is also related to the $D$ term. 

Another quantity of interest is the trace of the EMT $T^{\mu}_{\;\;\mu}$, which is related to the anomalous mass in the proton. The trace density is, 
\begin{equation}
\Theta(r_\perp) = M \int \frac{\dd^2 q_\perp}{(2\pi)^2} e^{-i\vec q_\perp \cdot \vec r_\perp} \Big\{ A(-q_\perp^2) + \frac{q_\perp^2}{4M^2} \big[A(-q_\perp^2) + 3 D(-q_\perp^2)\big] \Big\} = e(r_\perp) - 3 p(r_\perp)
\end{equation}
As a comparison, the Fourier transform of $t^{+-}$ in the Breit frame is, 
\begin{equation}
\frac{1}{2M} \mathcal T^{+-}(r_\perp)= e(r_\perp) - \frac{3}{2} p(r_\perp).
\end{equation}
The trace of the light-front EMT gives rise to a spurious contribution, 
\begin{equation}
\mathrm{tr} \, t(r_\perp) = t^{+-} - t^{11} - t^{22} = \Theta(r_\perp) - 2 \int \frac{\dd^2 q_\perp}{(2\pi)^2} e^{-i\vec q_\perp \cdot \vec r_\perp} q_\perp^2S_2(-q_\perp^2)
\end{equation}
Note that our definition here differs from the empirical definitions introduced by Polyakov et al. by a Darwin factor \cite{Polyakov:2018zvc}.

Energy conservation requires that the energy density integrated over the entire space gives the total mass of the system,
\begin{equation}
\int \dd^2 r_\perp \, e(r_\perp) = M. 
\end{equation}
This condition is fulfilled if 
\begin{equation}
A(0) = 1, \quad \lim_{q\to 0} q^2 D(q^2) = 0.
\end{equation} 
We will prove this is the case in our model. 

The state vector is an eigenstate of the light-front longitudinal momentum operator $\mathcal P^+$, 
\begin{equation}
\mathcal P^+ |\psi(p)\rangle = p^+ |\psi(p)\rangle.
\end{equation}
which is related to $T^{++}$ as,
\begin{equation}
\mathcal P^+ = \int\dd^3x \, T^{++}(x),
\end{equation}
where $\dd^3 x = (1/2)\dd x^- \dd^2 x_\perp$. Combining these two expressions we obtain, 
\begin{equation}
\lim_{q \to 0} t^{++}  = 2(P^+)^2
\end{equation}
which is consistent with our result Eq.~(\ref{eqn:forward_t++}). Combining this with Eq.~(\ref{eqn:GFF_A}), we obtain 
\begin{equation}\label{eqn:A(0)}
A(0) = 1.
\end{equation} 
Similar analysis is also applied to $T^{+i}$. Therefore, the momentum conservation guarantees the proper normalization of the $A$ term, independent of the truncation and other approximations that we employ. 

The proof of the second condition\footnote{Note that $D(q^2)$ is not necessarily finite in the forward limit $q \to 0$, which is actually the case for long-range interactions \cite{Varma:2020crx}. } $\lim_{q\to 0} q^2 D(q^2) = 0$ requires a consistency between the EMT and the Hamiltonian dynamics. 
The state vector satisfies the light-front Schrödinger equation (\ref{eqn:LF_Schrödinger_equation}),
and the light-front Hamiltonian operator $\mathcal P^-$ is related to $T^{+-}$ by (\ref{eqn:LF_Hamiltonian}).
Sandwiching $\mathcal P^-$ with the eigenstates and applying Eqs.~(\ref{eqn:LF_Schrödinger_equation}) \& (\ref{eqn:LF_Hamiltonian}), we obtain,
\begin{equation}\label{eqn:forward}
\langle \psi(p) |T^{+-}(0) |\psi(p)\rangle = 2(p_\perp^2+m^2).
\end{equation}
Indeed, this is exactly our result Eq.~(\ref{eqn:forward_t+-}) from the diagrams.
This expression together with Eqs.~(\ref{eqn:t+-} \& \ref{eqn:A(0)}) implies that, 
\begin{equation}\label{eqn:D_pole}
\lim_{q_\perp \to 0} q_\perp^2 D(-q_\perp^2) = 0.
\end{equation}
This condition is also related to the force balance inside the composite particles, known as the von Laue condition \cite{Laue:1911lrk}, which requires that the pressure integrated over the entire space vanishes,  
\begin{equation}
\int \dd^2 r_\perp \, p(r_\perp) = 0. 
\end{equation}

\subsection{Numerical results}\label{sec:numerics}

\begin{figure}[h]
\centering
\includegraphics[width=0.48\textwidth]{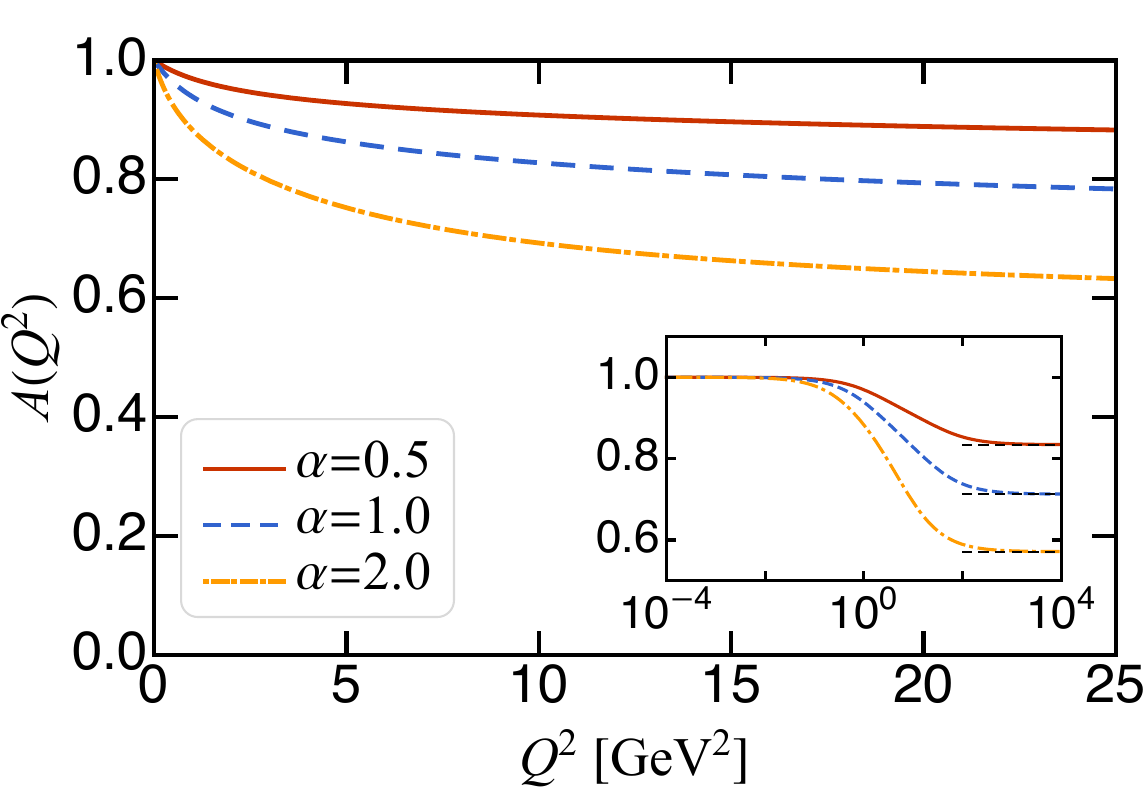}
\includegraphics[width=0.485\textwidth]{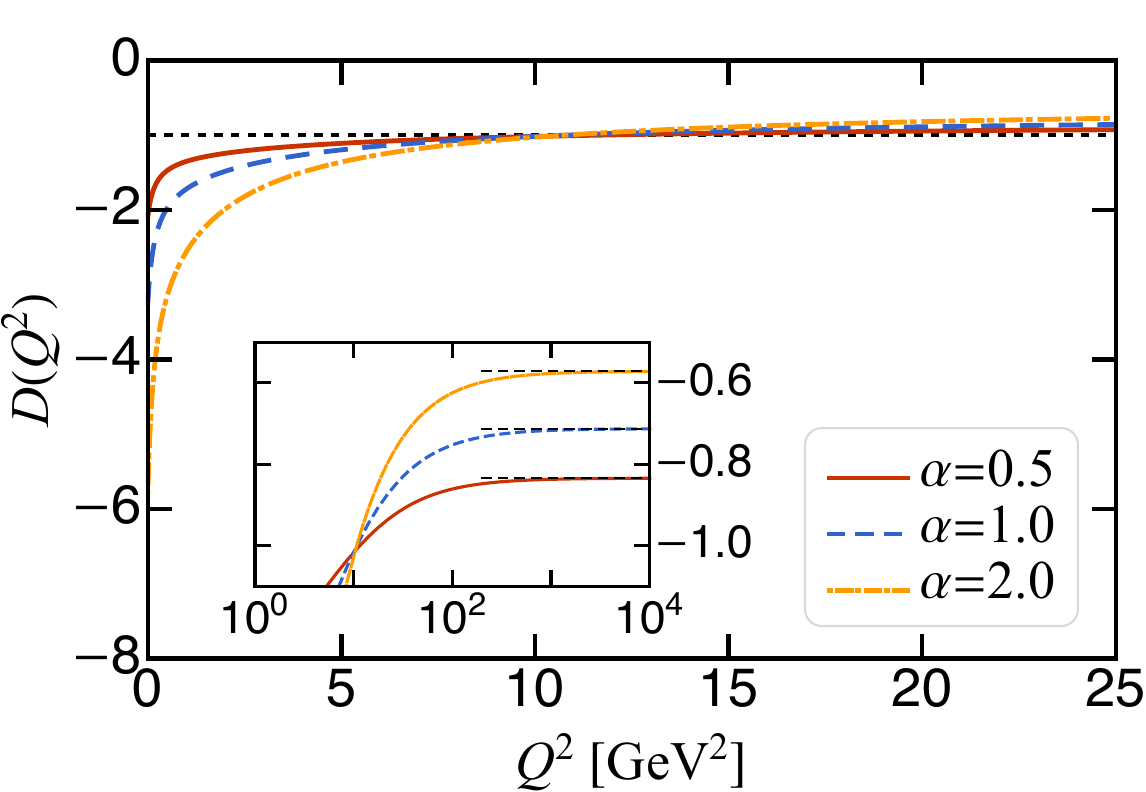}
\caption{The gravitational form factors $A(Q^2)$ and $D(Q^2)$ at various couplings. Here, $Q^2 = -q^2 = q_\perp^2$, $\alpha = g^2/(16\pi m^2)$. In the limit $Q \to \infty$, $A(Q^2)$ and $D(Q^2)$ approach to $Z$ and $-Z$, respectively, as indicated by the dashed lines in the inset panels for each coupling, where $Z$ is the field strength normalization constant. }
\label{fig:A_and_D}
\end{figure}

\begin{figure}[h]
\centering
\includegraphics[width=0.5\textwidth]{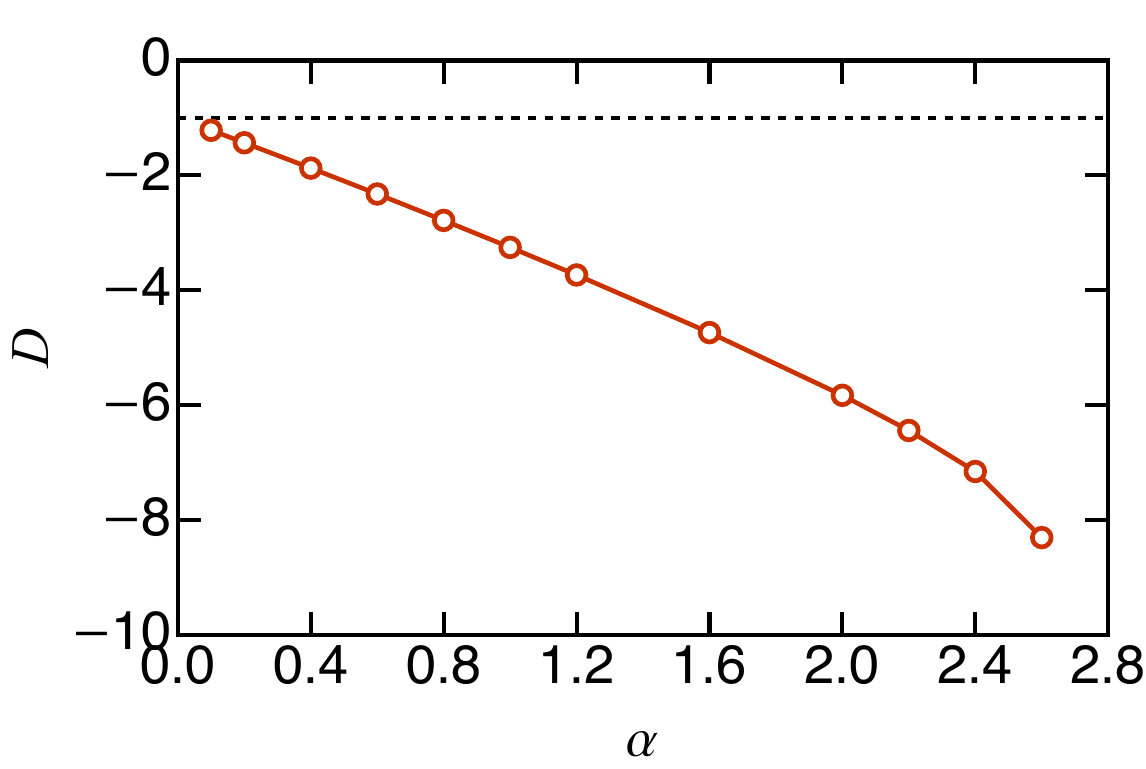}
\caption{The $D$ term $D = D(0)$ as a function of the coupling $\alpha = g^2/(16\pi m^2)$.}
\label{fig:D0}
\end{figure}

The numerical results of the GFFs are shown in Fig.~\ref{fig:A_and_D}. In this figure, we compare results at various couplings from the perturbative regime $\alpha = 0.5$ to the strong coupling regime $\alpha = 2.0$, where the dimensionless coupling constant is related to the physical coupling $g$ as $\alpha = g^2/(16\pi m^2)$. In this figure, the $A$ term is normalized to unity in the forward limit, viz., $A(0) = 1$, and the $D$ term in the same limit is finite and negative $D = D(0) < 0$. In fact, $D<-1$ in our model. Fig.~\ref{fig:D0} shows the $D$ as a function of the coupling $\alpha$. As the coupling increases, $D$ becomes more negative. The value of $D$ appears quite sensitive to the coupling, thus providing a good probe of the interaction strength of the system. 
At large $Q^2$, the form factors approach to their one-body contribution, i.e., $A(Q^2\to \infty) = A_1 = Z$, and $D(Q^2\to\infty) = D_1 = -Z$, where $Z$ is the field strength renormalization constant. 

\begin{figure}
\centering
\includegraphics[width=0.45\textwidth]{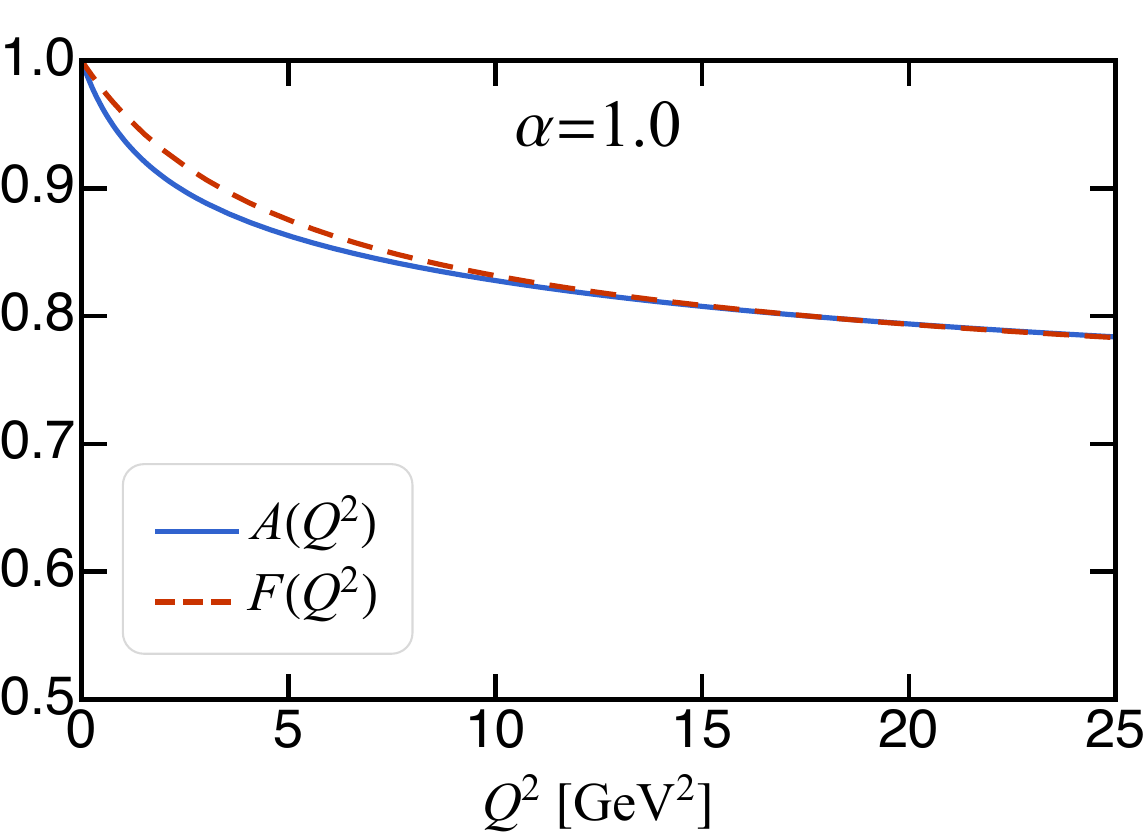}\quad 
\raisebox{-0.015\height}{\includegraphics[width=0.455\textwidth]{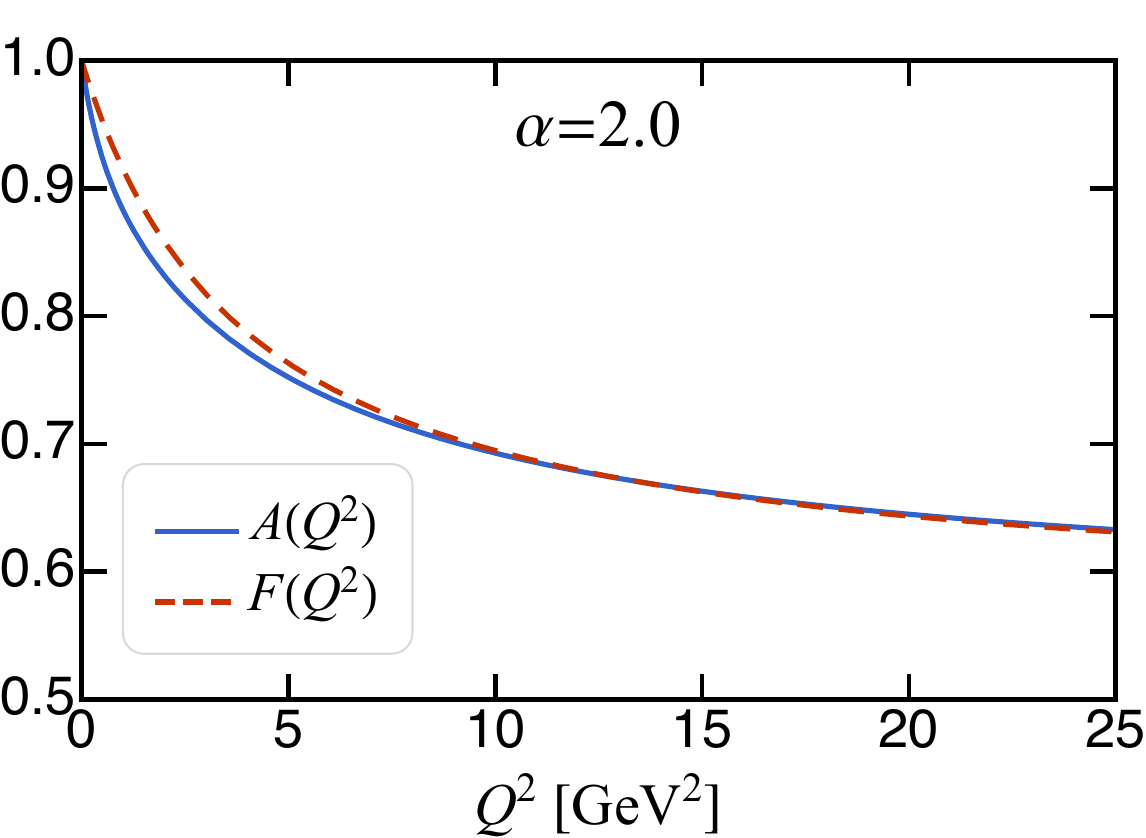}}
\caption{The gravitational form factor $A(Q^2)$ as compared with the charge form factor $F(Q^2)$ at $\alpha = 1.0$ and $\alpha = 2.0$. Here, $Q^2 = -q^2 = q_\perp^2$, $\alpha = g^2/(16\pi m^2)$.}
\label{fig:A_and_F}
\end{figure}

\begin{figure}
\centering
\includegraphics[width=0.5\textwidth]{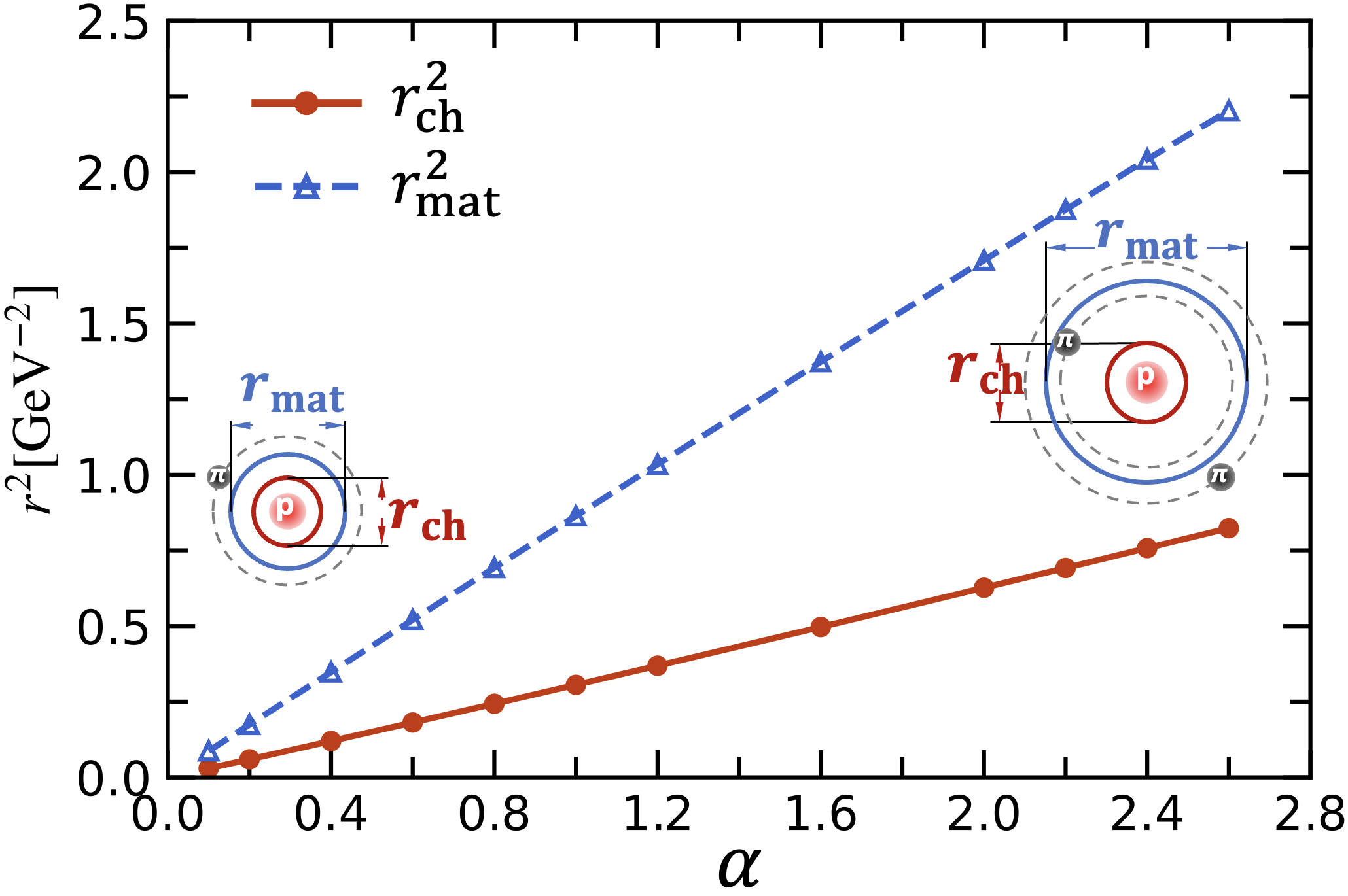}
\caption{Comparison of the matter radius and the charge radius as functions of the coupling. As the coupling increases, the two-pion sea contribution increases which leads to a dramatic increase of the matter radius of the system. }
\label{fig:radii}
\end{figure}

\begin{figure}
\centering
\includegraphics[width=0.6\textwidth]{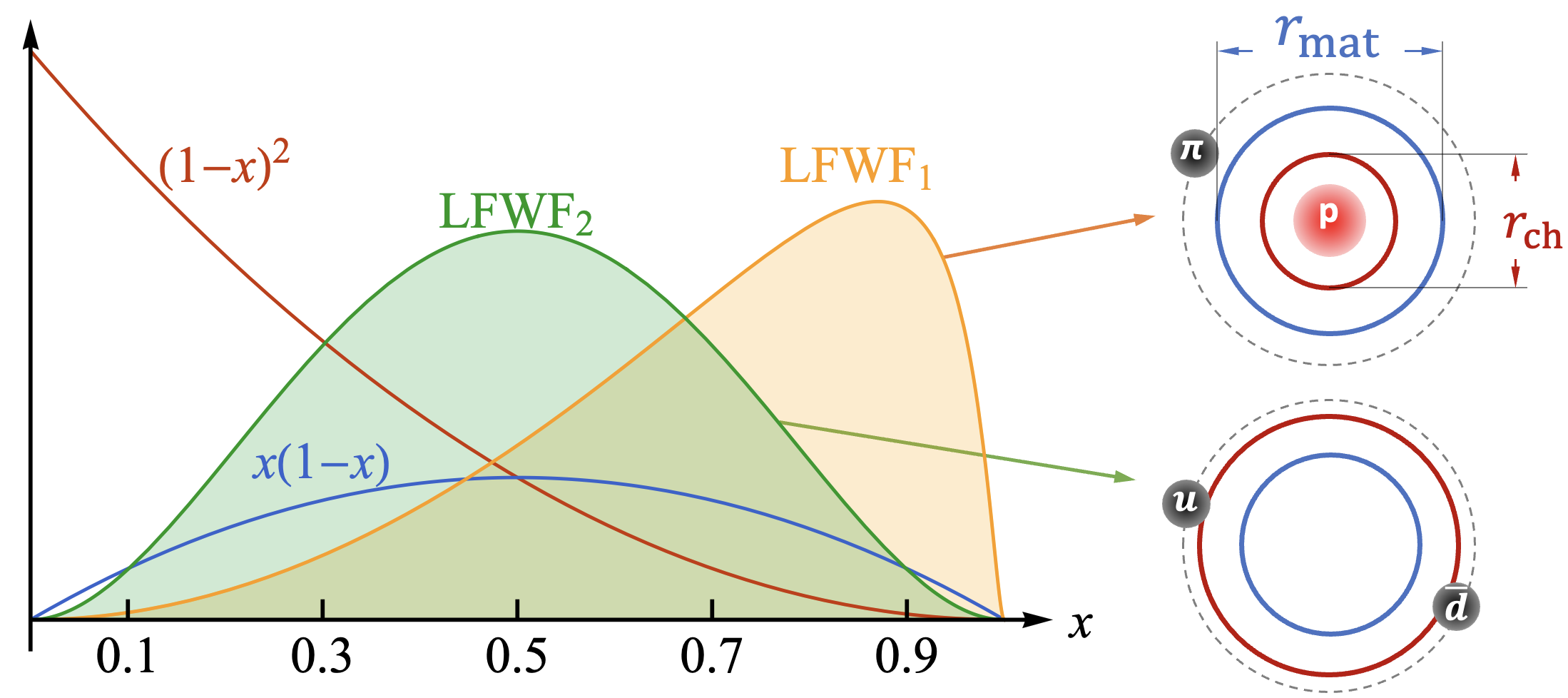}
\caption{Comparison of the matter radius and the charge radius for two systems.}
\label{fig:radii2}
\end{figure}

Figure~\ref{fig:A_and_F} compares the GFF $A(Q^2)$ and the charge form factor $F(Q^2)$. Both quantities approach to the same limit at large $Q^2$. However, at small $Q^2$, $A(Q^2)$ appears softer than $F(Q^2)$, which implies a larger matter radius than the charge radius, viz., $r^2_\text{mat} > r^2_\text{ch}$, where $r^2_\text{mat} = -6A'(Q^2 =0)$ and $r^2_\text{ch} = -6F'(Q^2)$. 
Indeed, $r^2_\text{mat}$ is always larger than $r^2_\text{ch}$ for the couplings we consider as shown in Fig.~\ref{fig:radii}. 
Recall that for mesons, the relative size is reversed \cite{Li:2016wwu}. These can be understood in the LFWF representation (see Sect.~\ref{sec:LFWF_representation} for more details). In this representation, the charge radius $r^2_\text{ch} = (3/2) \langle e_q (1-x)^2 r^2_\perp \rangle$ while the matter radius $r^2_\text{mat} = (3/2) \langle x(1-x) r^2_\perp \rangle$. The cartoon in Fig.~\ref{fig:radii2} illustrates the origins of the mean radii for two systems. For the pion, the LFWF is approximately symmetric with respect to $x$. This shows its charge radius is larger than its matter radius. By contrast, for our pion-dressed proton, the LFWF concentrates on the $x_p \sim 1$ side, and its matter radius becomes larger than its charge radius. 

\begin{figure}
\centering
\includegraphics[width=0.45\textwidth]{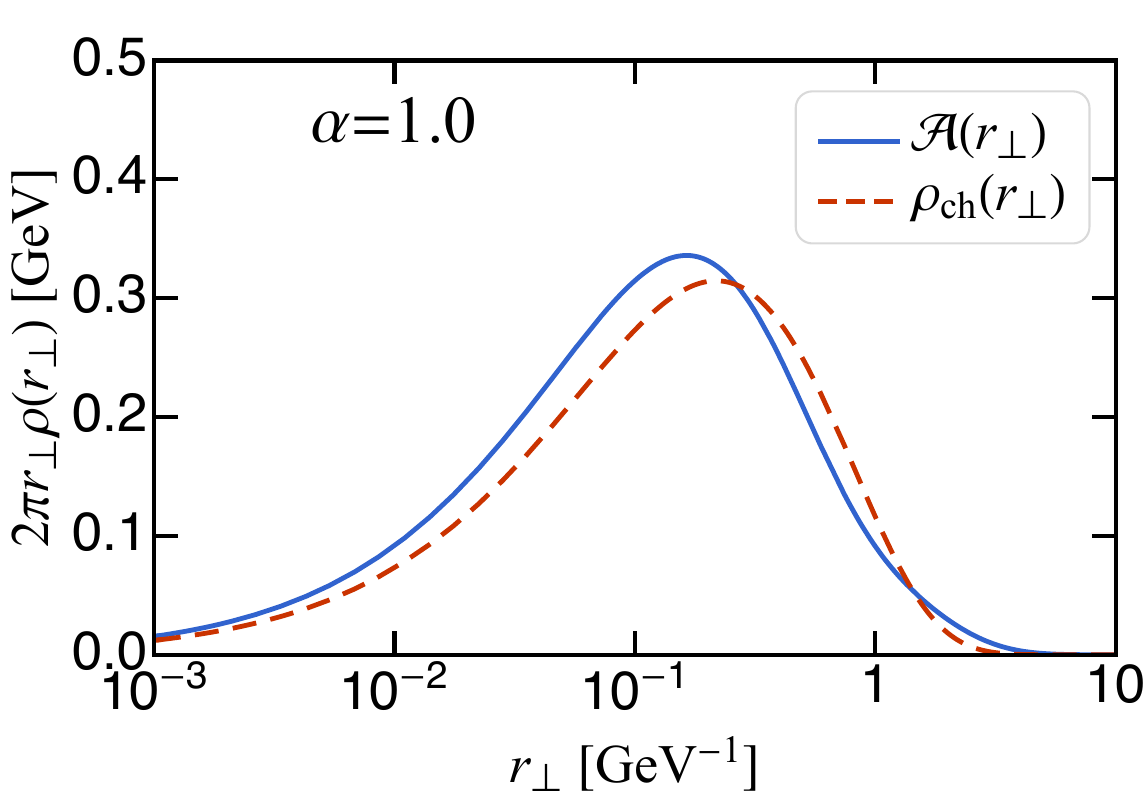}\quad
\includegraphics[width=0.45\textwidth]{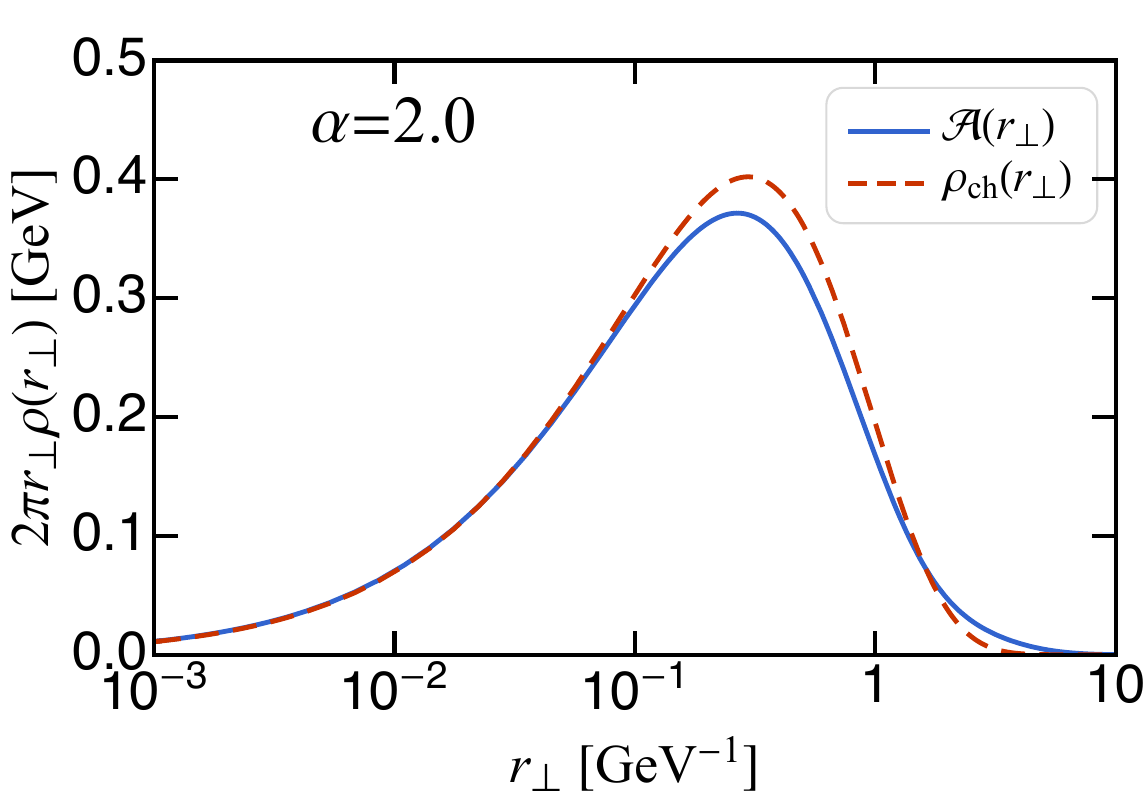}
\caption{The transverse matter density $\mathcal A(r_\perp)$ as compared with the transverse charge density $\rho_\text{ch}(r_\perp)$ at $\alpha = 1.0$ and $\alpha = 2.0$, where the dimensionless coupling $\alpha = g^2/(16\pi m^2)$.}
\label{fig:Adensity}
\end{figure}

\begin{figure}
\centering
\includegraphics[width=0.45\textwidth]{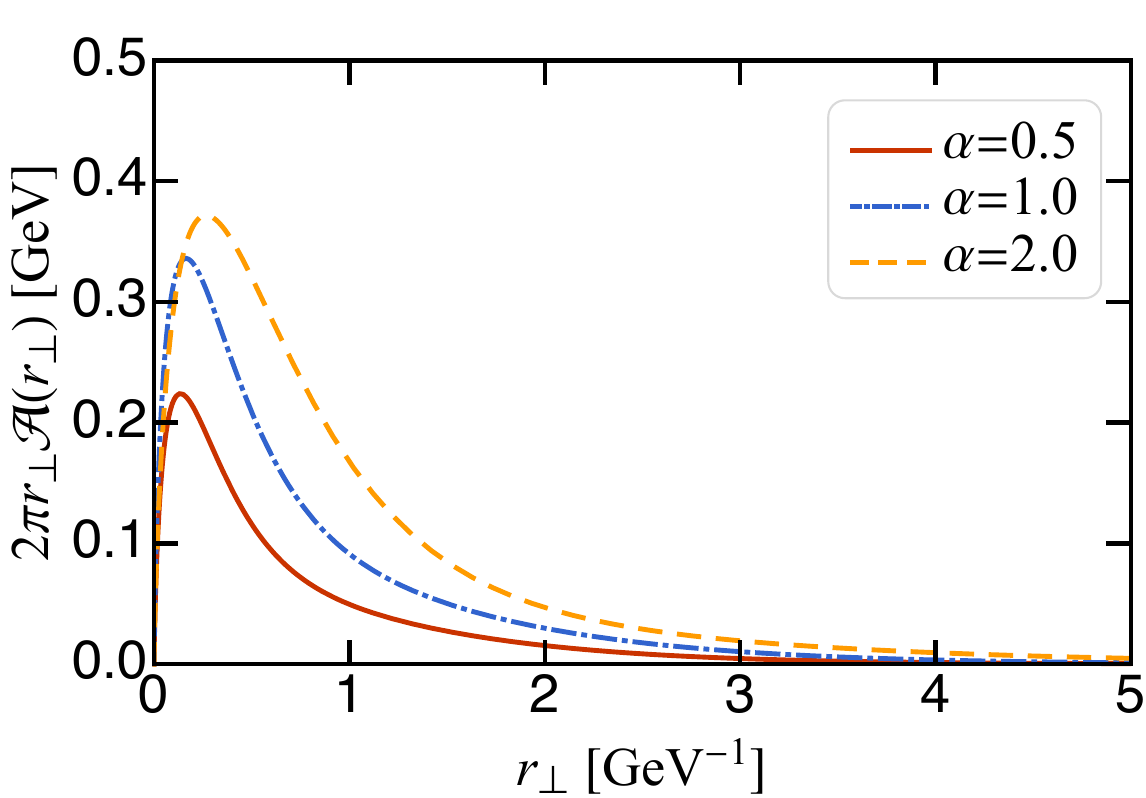}\quad
\raisebox{-0.01\height}{\includegraphics[width=0.47\textwidth]{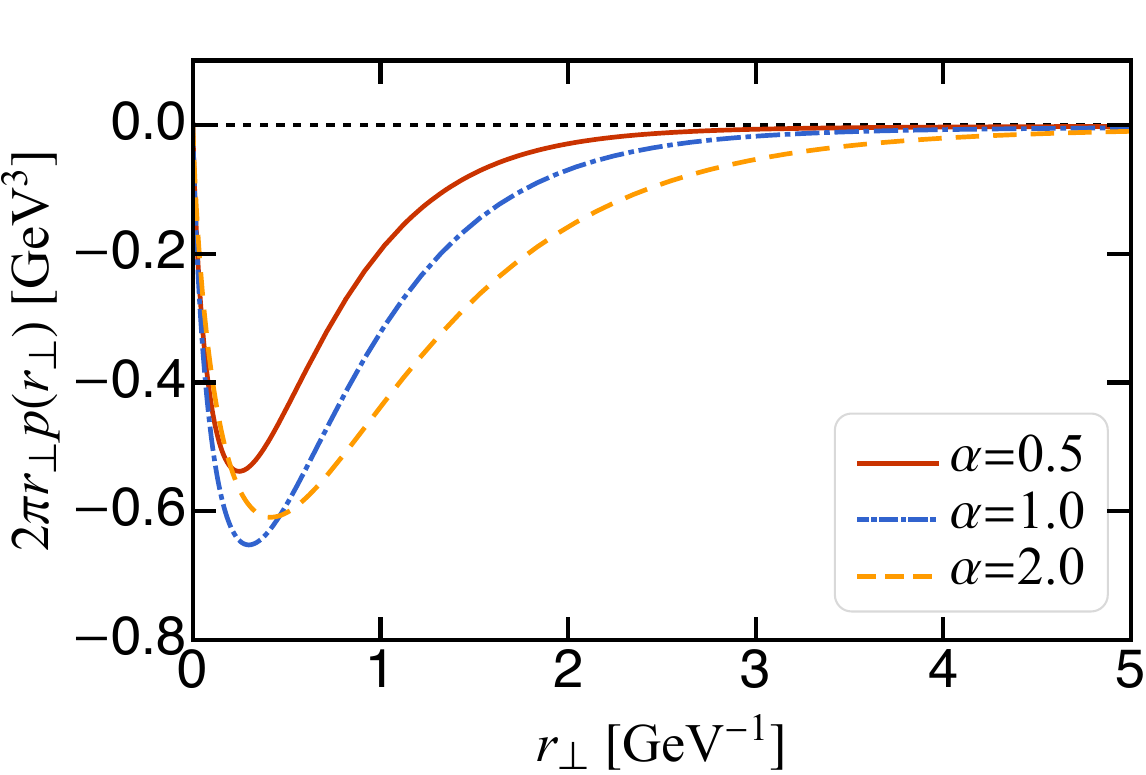}}
\caption{The transverse matter density $\mathcal A(r_\perp)$ and pressure $p(r_\perp)$ at selected couplings.}
\label{fig:A_and_pressure}
\end{figure}

From the form factors, we can extract the corresponding transverse densities by a Fourier transformation. To avoid the numerical difficulties, we first fit the form factors with the multi-monopole function, 
\begin{equation}
F(Q^2) = Z + \frac{(1-Z)a_1}{1 + {Q^2}/{\Lambda_1^2}} + \frac{(1-Z)(1-a_1)}{1 + {Q^2}/{\Lambda_2^2}}.
\end{equation}
The corresponding density is, 
\begin{equation}
\rho(r_\perp) = Z\delta^2(r_\perp) + \frac{(1-Z)}{2\pi} a_1 \Lambda_1^2 K_0(\Lambda_1 r_\perp) 
+ \frac{(1-Z)}{2\pi} (1-a_1) \Lambda_2^2 K_0(\Lambda_2 r_\perp).
\end{equation}
Chiral effective field theory predicts a pion cloud $\sim\exp(-2M_\pi r_\perp)$ at the periphery of the nucleon transverse charge density ($r_\perp = O(M^{-1}_\pi)$) \cite{Granados:2013moa, Granados:2015lxa, Granados:2015rra}. The ansatz we adopt here is qualitatively in agreement with this picture. The extracted matter density\footnote{This quantity is also called the mass density, longitudinal momentum density, the tensor density or simply the $A$-density in the literature. However, from our analysis above, $\mathcal A(r_\perp)$ is the density associated with all three momenta $P^+, \vec P_\perp$. On the other hand, both the proper energy density $e(r_\perp)$ and the invariant mass squared density $\mathcal E(r_\perp)$ (Sect.~\ref{sec:t+-}) differ from $\mathcal A(r_\perp)$. For this reason, we feel it is more justified to call this quantity the matter density. } $\mathcal A(r_\perp)$ is shown in Fig.~\ref{fig:Adensity}, in comparison with the charge density $\rho_\text{ch}(r_\perp)$, the Fourier transform of the charge form factor $F(Q^2)$. Fig.~\ref{fig:A_and_pressure} shows the transverse matter density $\mathcal A(r_\perp)$ and pressure $p(r_\perp)$ for various couplings. The expected node within the pressure is squeezed to the origin $r_\perp = 0$ due to the point-like repulsive core $\propto \delta^2(r_\perp)$ (not shown in the figure). 

\section{Light-front wave function representation}\label{sec:LFWF_representation}

In this section, we further analyze the LFWF representation. Our goal is to obtain a general non-perturbative representation independent of the interactions.

\subsection{$t^{++}$ and the $A$ term}

We summarize the hadron matrix element $t^{++}$ computed in Sect.~\ref{sec:HMEs} as follows, 
\begin{align}
t^{++}  =\,&  2(P^+)^2 Z \\
+\,& 2(P^+)^2\int \frac{\dd x}{2x (1-x)}\int \frac{\dd^2k_\perp}{(2\pi)^3} \psi_2(x, \vec k_\perp) \psi_2^*(x, \vec k_\perp-x\vec q_\perp) (1-x) \\
+\,& 2(P^+)^2\int \frac{\dd x}{2x (1-x)}\int \frac{\dd^2k_\perp}{(2\pi)^3} \psi_2(x, \vec k_\perp) \psi_2^*(x, \vec k_\perp+(1-x)\vec q_\perp) x\\
+\,& 2(P^+)^2 \frac{1}{2!} \int \frac{\dd x}{2x}\int \frac{\dd^2 k_\perp}{(2\pi)^3}\int \frac{\dd x'}{2x'(1-x-x')}
\int \frac{\dd^2 k'_\perp}{(2\pi)^3} \\
&\times \psi_3(x, \vec k_\perp, x', \vec k'_\perp) \psi_3^*(x, \vec k_\perp - x\vec q_\perp, x', \vec k'_\perp - x'\vec q_\perp) 
(1-x-x')  \nonumber \\
+\,&  2(P^+)^2\frac{1}{2!}\int \frac{\dd x}{2x}\int \frac{\dd^2 k_\perp}{(2\pi)^3}\int\frac{\dd x'}{2x'(1-x-x')} \int \frac{\dd^2k'_\perp}{(2\pi)^3} \\
& \times \psi_3(x, \vec k_\perp, x', \vec k'_\perp) \psi_3^*(x, \vec k_\perp + (1- x) \vec q_\perp, x', \vec k'_\perp -x' \vec q_\perp) x \nonumber \\
+\,& 2(P^+)^2\frac{1}{2!}\int \frac{\dd x}{2x}\int \frac{\dd^2 k_\perp}{(2\pi)^3}\int\frac{\dd x'}{2x'(1-x-x')} \int \frac{\dd^2k'_\perp}{(2\pi)^3} \nonumber\\
& \times \psi_3(x, \vec k_\perp, x', \vec k'_\perp) \psi_3^*(x, \vec k_\perp -  x\vec q_\perp, x', \vec k'_\perp +(1-x') \vec q_\perp) x' 
\end{align}
Based on the above expression, it is not hard to conclude a general formula for $t^{++}$:
\begin{equation}\label{eqn:off-forward_++}
t^{++} = 2(P^+)^2 \sum_n \int \big[\dd x_i \dd^2 k_{i\perp} \big]_n \sum_j x_j \psi_n(\{x_i, \vec k_{i\perp}\})  \psi_n(\{x_i, \vec k_{i,j\perp}\}) 
\end{equation}
where, 
\begin{equation}
\vec k_{i,j\perp} = \begin{cases}
\vec k_{i\perp} - x_i \vec q_\perp, &  \text{spectator: }i\ne j \\
\vec k_{i\perp} + (1- x_i) \vec q_\perp, & \text{struck parton: } i = j\\
\end{cases}
\end{equation}
And the $n$-body integration measure is defined as, 
\begin{equation}
\int\big[\dd x_i \dd^2 k_{i\perp}\big]_n = \frac{1}{S_n}\prod_{i=1}^{n} \int \frac{\dd x_i}{2x_i} 2\delta(\sum_i x_i -1) \int \frac{\dd^2k_{i\perp}}{(2\pi)^3} (2\pi)^3\delta^2(\sum_i k_{i\perp})
\end{equation}
where, $S_n$ is the symmetry factor. 

Using the transverse coordinate representation introduced in Appendix~\ref{sec:coordinate_reprensentation}, the hadron matrix element (\ref{eqn:off-forward_++}) can  be written as, 
\begin{equation}\label{eqn:off-forward_++_coord}
t^{++} = 2(P^+)^2 \sum_n \int \big[\dd x_i \dd^2 r_{i\perp}\big]_n  \Big| \widetilde\psi_n(\{x_i, \vec r_{i\perp}\})\Big|^2 \sum_j x_j e^{i \vec r_{j\perp}\cdot \vec q_\perp}
\end{equation}
The corresponding GFF $A$ is, 
\begin{equation}\label{eqn:A_coord}
A(-q_\perp^2) =\sum_n \int \big[\dd x_i \dd^2 r_{i\perp}\big]_n  \Big| \widetilde\psi_n(\{x_i, \vec r_{i\perp}\})\Big|^2 \sum_j x_j e^{i \vec r_{j\perp}\cdot \vec q_\perp}
\end{equation}
The light-front distribution is defined as the Fourier transform of the form factor $A$, 
\begin{align}\label{eqn:density_A}
\mathcal A(r_\perp) =\,& \int \frac{\dd^2 q_\perp}{(2\pi)^2} e^{-i\vec q_\perp \cdot \vec r_\perp}A(-q_\perp^2) \\
=\,& \sum_n \int \big[\dd x_i \dd^2 r_{i\perp}\big]_n  \Big| \widetilde\psi_n(\{x_i, \vec r_{i\perp}\})\Big|^2 \sum_j x_j \delta^2(\vec r_\perp - \vec r_{j\perp})
\end{align}
This expression is in agreement with the classic results by Brodsky et al. \cite{Brodsky:2000ii}. It generalizes the one-body density on the light front.  Hence $\mathcal A(r_\perp)$ should be understood as the one-body (number) density. 

The matter radius $r_\text{mat}^2$ is defined as the slope of the form factor $A(q^2)$ multiplied by 6, which is equivalent to $3/2$ times of the mean transverse squared radius,
\begin{equation}
r^2_\text{mat} = \frac{6}{A(0)}\frac{\dd }{\dd q^2}A(q^2)\Big|_{q^2 \to 0} = \frac{3}{2} \int \dd^2 r_\perp \, r^2_\perp \mathcal A(r_\perp). 
\end{equation}
In LFWF representation, 
\begin{equation}
r^2_\text{mat} = \frac{3}{2} \sum_n \int \big[\dd x_i \dd^2 r_{i\perp}\big]_n  \Big| \widetilde\psi_n(\{x_i, \vec r_{i\perp}\})\Big|^2 \sum_j x_j r_{j\perp}^2.
\end{equation}

\subsection{$t^{+-}$ and the $D$ term}\label{sec:t+-}

We summarize the hadron matrix element $t^{+-}$ computed in Sect.~\ref{sec:HMEs} as follows.
The one-body contribution (diagram $a$, $b$) is, 
\begin{equation}
t^{+-}_1 = Z \big[ 2(m^2 + P^2_\perp) - \frac{1}{2} q_\perp^2 \big] 
\end{equation}
The two-body contribution (diagram $c$, $d$, $\bar b$, $f$) reads,
\begin{align}
t^{+-}_2 =\,& 
\int\frac{\dd x}{2x(1-x)} \int \frac{\dd^2\ell_\perp}{(2\pi)^3} 
\bigg\{ \nonumber \\
& \psi_2(x, \vec\ell_\perp+\frac{1}{2}x\vec q_\perp)
\psi_2^*(x, \vec\ell_\perp-\frac{1}{2}x\vec q_\perp)
\frac{\big[2(\vec \ell_\perp-(1-x)\vec P_\perp)^2 + 2m^2 - \frac{1}{2} q_\perp^2\big]}{1-x} \\
+\,& \psi_2(x, \vec\ell_\perp-\frac{1}{2}(1-x)\vec q_\perp)
\psi_2^*(x, \vec\ell_\perp+\frac{1}{2}(1-x)\vec q_\perp)
\frac{\big[2(\vec \ell_\perp+x\vec P_\perp)^2 + 2\mu^2 - \frac{1}{2} q_\perp^2\big]}{x} \\
-& 2\psi_2(x, \vec \ell_\perp)  \psi^*_2(x, \vec \ell_\perp - x\vec q_\perp) 
 \Big(\frac{\ell_\perp^2+\mu^2}{x} + \frac{\ell_\perp^2+m^2}{1-x}-M^2\Big)\bigg\}
\end{align}
Similarly, the three-body contribution (diagram $\bar f$, $g$, $h$) reads, 
\begin{align}
t^{+-}_3 =\,& 
\frac{1}{2!} \int\frac{\dd x}{2x} \int \frac{\dd^2\ell_\perp}{(2\pi)^3} 
\int\frac{\dd x'}{2x'(1-x-x')} \int \frac{\dd^2\ell'_\perp}{(2\pi)^3} 
\bigg\{ \nonumber \\
& \psi_3(x, \vec\ell_\perp+\frac{1}{2}x\vec q_\perp, x', \vec\ell'_\perp+\frac{1}{2}x'\vec q_\perp)
\psi^*_3(x, \vec\ell_\perp-\frac{1}{2}x\vec q_\perp, x', \vec\ell'_\perp-\frac{1}{2}x'\vec q_\perp) \nonumber \\
&\quad\times\frac{2(\vec\ell_\perp+\vec\ell'_\perp-(1-x-x')\vec P_\perp)^2+2m^2-\frac{1}{2}\vec q^2_\perp}{1-x-x'} \\
+\, &  \psi_3(x, \vec\ell_\perp-\frac{1}{2}(1-x)\vec q_\perp, x', \vec\ell'_\perp+\frac{1}{2}x'\vec q_\perp)
\psi^*_3(x, \vec\ell_\perp+\frac{1}{2}(1-x)\vec q_\perp, x', \vec\ell'_\perp-\frac{1}{2}x'\vec q_\perp) \nonumber \\
&\quad\times \frac{2(\vec\ell_\perp+x\vec P_\perp)^2+2\mu^2-\frac{1}{2}\vec q^2_\perp}{x} \\
-\,& 2 \psi_3(x, \vec \ell_\perp, x', \vec \ell'_\perp) 
 \psi^*_3(x, \vec \ell_\perp - x\vec q_\perp, x', \vec \ell'_\perp - x'\vec q_\perp) \nonumber \\
& \times\Big[ \frac{(\vec \ell_\perp + \vec \ell'_\perp)^2+m^2}{1-x-x'} + \frac{\ell_\perp^2+\mu^2}{x} +  \frac{\ell'^2_\perp+\mu^2}{x'} -M^2\Big]\bigg\}
\end{align}

From the above expressions, the general $n$-body contribution should be,
\begin{multline}\label{eqn:off-forward_t+-}
t^{+-}_n = 2\int \big[\dd x_i \dd^2 k_{i\perp} \big]_n \sum_j \psi_n^*(\{x_i, \vec k_{i,j\perp}^+\})
\psi_n(\{x_i, \vec k_{i,j\perp}^-\}) \frac{(\vec k_{j\perp}+x_j\vec P_\perp)^2+m_j^2-\frac{1}{4}q_\perp^2}{x_j} \\
+ 2\int \big[\dd x_i \dd^2 k_{i\perp} \big]_n \psi_n^*(\{x_i, \vec k_{i\perp}\})
\psi_n(\{x_i, \vec k_{i,n\perp}\}) \Big[M^2 - \sum_j \frac{\vec k_{j\perp}^2+m_j^2}{x_j}\Big]
\end{multline}
where, 
\begin{align}
\vec k_{i,n\perp} =\,& \begin{cases}
\vec k_{i\perp} - x_i \vec q_\perp, &  \text{pion, i.e. }i\ne n \\
\vec k_{i\perp} + (1- x_i) \vec q_\perp, & \text{nucleon, i.e. }i = n \\
\end{cases} \\
& \nonumber \\
\vec k^+_{i,j\perp} =\,& \begin{cases}
\vec k_{i\perp} + \frac{1}{2} x_i \vec q_\perp, &  \text{spectator: }i\ne j \\
\vec k_{i\perp} - \frac{1}{2}(1- x_i) \vec q_\perp, & \text{struck parton: } i = j\\
\end{cases} \\
& \nonumber \\
\vec k^-_{i,j\perp} =\,& \begin{cases}
\vec k_{i\perp} - \frac{1}{2} x_i \vec q_\perp, &  \text{spectator: }i\ne j \\
\vec k_{i\perp} + \frac{1}{2}(1- x_i) \vec q_\perp, & \text{struck parton: } i = j
\end{cases}
\end{align}
The first line of Eq.~(\ref{eqn:off-forward_t+-}) represents the off-forward kinetic energy whereas the second line is the off-forward potential energy (mass eigenvalue minus kinetic energy), thus generalizing Eq.~(\ref{eqn:T+-_int}) to the off-forward region, 
\begin{equation}\label{eqn:off-forward_T+-_int}
\langle \{x_i p^+, \vec k_{i\perp}+x_i (\vec p_\perp+\vec q_\perp)\}_n | T^{+-}_\text{int} (0) |\psi(p)\rangle 
= -2\Gamma_n(\{x_i, \vec k_{i,n\perp}\}).
\end{equation}

In the transverse coordinate-space, 
\begin{multline}
t^{+-}_n = 2\int \big[\dd x_i \dd^2 r_{i\perp} \big]_n \widetilde\psi_n^*(\{x_i, \vec r_{i\perp}\})
\sum_j  e^{i\vec r_{j\perp}\cdot\vec q_\perp} \Big( \frac{-\nabla_{j\perp}^2+m_j^2-\frac{1}{4}q_\perp^2}{x_j} + x_j\vec P_\perp^2\Big) 
 \widetilde\psi_n(\{x_i, \vec r_{i\perp}\}) \\
- 2\int \big[\dd x_i \dd^2 r_{i\perp} \big]_n \widetilde\psi_n^*(\{x_i, \vec r_{i\perp}\}) \Big[\sum_j \frac{-\nabla^2_{j\perp}+m_j^2}{x_j} - M^2\Big]
\widetilde\psi_n(\{x_i, \vec r_{i\perp}\}) e^{i \vec r_{n\perp} \cdot \vec q_\perp} 
\end{multline}
The corresponding total density is the one-body light-cone energy density 
\begin{equation}
\frac{1}{2}\mathcal T^{+-}(r_\perp) = \frac{1}{2}\sum_n \mathcal T_n^{+-}(r_\perp) \equiv \mathcal E(r_\perp) +  P_\perp^2 \mathcal A(r_\perp) = \mathcal T(r_\perp) + \mathcal V(r_\perp) +  P_\perp^2 \mathcal A(r_\perp),
\end{equation}
 where the one-body (off-forward) kinetic energy density is,
\begin{align}
\mathcal T(r_\perp) =\,& \sum_n \int \big[\dd x_i \dd^2 r_{i\perp} \big]_n  \widetilde\psi_n^*(\{x_i, \vec r_{i\perp}\}) 
\sum_j \delta^2(r_\perp - r_{j\perp}) \frac{-\vec\nabla^2_{j\perp} + m_j^2 - \frac{1}{4}\cev\nabla^2_\perp}{x_j} \widetilde\psi_n(\{x_i, \vec r_{i\perp}\}) \\
 =\,& \sum_n \int \big[\dd x_i \dd^2 r_{i\perp} \big]_n \sum_j \delta^2(r_\perp - r_{j\perp})
 \widetilde\psi_n^*(\{x_i, \vec r_{i\perp}\}) 
 \frac{-\frac{1}{4}\tensor\nabla^2_{j\perp} + m_j^2}{x_j} \widetilde\psi_n(\{x_i, \vec r_{i\perp}\}) 
 \end{align}
 The one-body potential energy density in our case only involves the diagonal Fock sector contributions, 
\begin{equation}
\mathcal V(r_\perp) = \sum_n \int \big[\dd x_i \dd^2 r_{i\perp} \big]_n  \widetilde\psi_n^*(\{x_i, \vec r_{i\perp}\}) 
\delta^2(r_\perp - r_{n\perp}) \big( M^2 - s_n \big)  \widetilde\psi_n(\{x_i, \vec r_{i\perp}\})
\end{equation}
where, $s_n = \sum_i (-\vec\nabla^2_{i\perp} + m_i^2)/x_i$ is the $n$-body light-cone kinetic energy. 
Note that the Dirac-$\delta$ only samples the $n$-th parton, the mock nucleon. This is because of the quenched approximation: all interaction is associated with the mock nucleon. It is remarkable that the EMT ``knows" the quenched approximation. 

The Fourier transforms of $\mathcal E(r_\perp)$, $\mathcal T(r_\perp)$ and $\mathcal V(r_\perp)$ are the corresponding form factors, $E(-q_\perp^2)$, $T(-q_\perp^2)$ and $V(-q_\perp^2)$. At zero-momentum transfer, the squared invariant mass form factor gives the total squared invariant mass $E(0) = T(0) + V(0) = M^2$. 
From these expressions, we obtain, 
\begin{equation}
q_\perp^2 D(-q_\perp^2)  =  2 E(-q^2_\perp) - 2\big[E(0) + \frac{1}{4}q^2_\perp\big] A(-q_\perp^2)
\end{equation}
The von Laue mechanical equilibrium condition is automatically fulfilled as long as $A(0) = 1$, a consequence of  momentum conservation~(\ref{eqn:A(0)}). 
The GFF $D(q^2)$ is,
\begin{multline}
D(-q_\perp^2) = 2\sum_n \int \big[\dd x_i \dd^2 r_{i\perp} \big]_n  \widetilde\psi_n^*(\{x_i, \vec r_{i\perp}\})  \\
\times \sum_j  \Big\{ \frac{e^{i\vec r_{j\perp}\cdot\vec q_\perp}-e^{i\vec r_{n\perp}\cdot\vec q_\perp}}{q_\perp^2}  \frac{-\nabla_{j\perp}^2+m_j^2-x_j^2M^2}{x_j}  - \frac{1+x_j^2}{4x_j} e^{i\vec r_{j\perp}\cdot\vec q_\perp} \Big\}
 \widetilde\psi_n(\{x_i, \vec r_{i\perp}\}). 
 \end{multline}
In particular, the $D$ term is finite, 
\begin{multline}
D \equiv D(0) =  -1 + 2\sum_n \int \big[\dd x_i \dd^2 r_{i\perp} \big]_n  \widetilde\psi_n^*(\{x_i, \vec r_{i\perp}\})  \\
\times \sum_j  \frac{1}{x_j}\Big\{ (r_{n\perp}^2 - r^2_{j\perp})(-\nabla_{j\perp}^2+m_j^2-x_j^2M^2)  + \mathsmaller{\frac{1}{4}}(x_j^2-1) \Big\}  \widetilde\psi_n(\{x_i, \vec r_{i\perp}\}).
 \end{multline}
The $D$ term contains a term $\sum_j r^2_{j\perp} p^2_{j\perp} $ resembling the virial term. This term measures how the wave function scales in terms of a dilation transformation in the transverse direction. 
The Fourier transform of the $D$ term is a quantity of interest. 
\begin{multline}
\mathcal D(r_\perp) = 2\sum_n \int \big[\dd x_i \dd^2 r_{i\perp} \big]_n  \widetilde\psi_n^*(\{x_i, \vec r_{i\perp}\})  \\
\times \sum_j  \Big\{\ln \frac{\big|\vec r_\perp - \vec r_{n\perp}\big|}{\big|\vec r_\perp - \vec r_{j\perp}\big|} \frac{-\nabla_{j\perp}^2+m_j^2-x_j^2M^2}{x_j}  - \frac{1+x_j^2}{4x_j} \delta^2(\vec r_\perp - \vec r_{j\perp}) \Big\}
 \widetilde\psi_n(\{x_i, \vec r_{i\perp}\}). 
 \end{multline}

\begin{figure}
\centering
\includegraphics[width=0.45\textwidth]{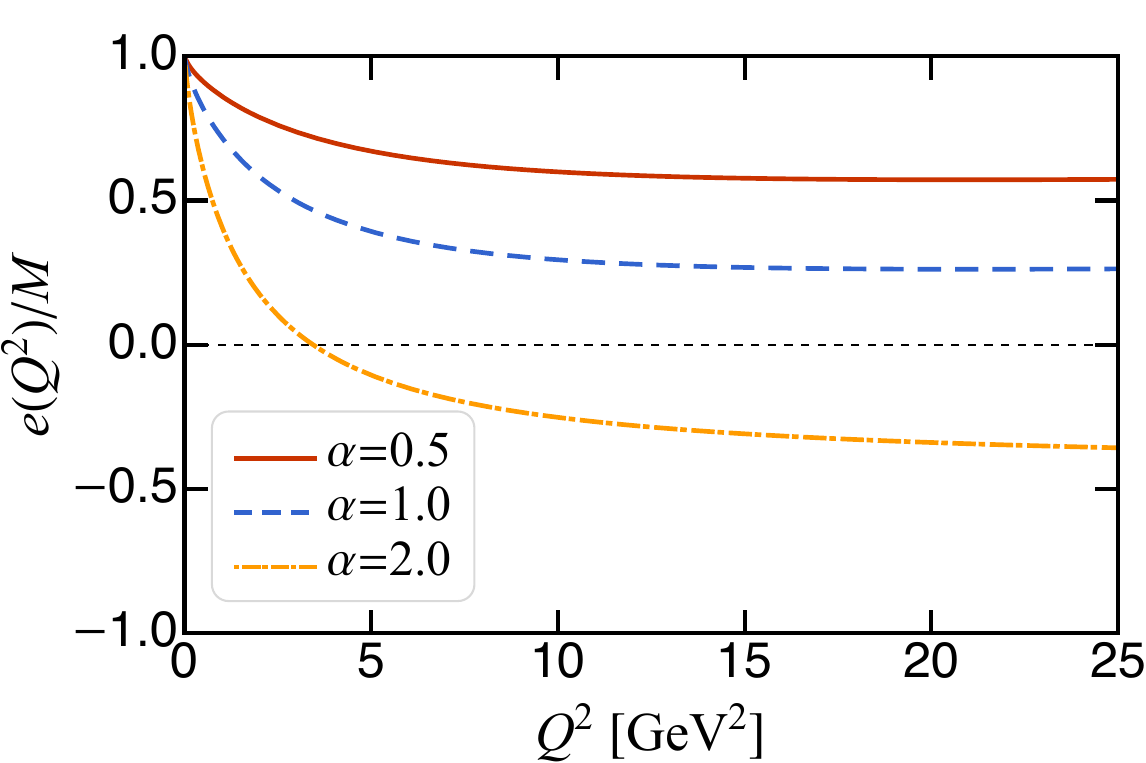}\quad
\includegraphics[width=0.44\textwidth]{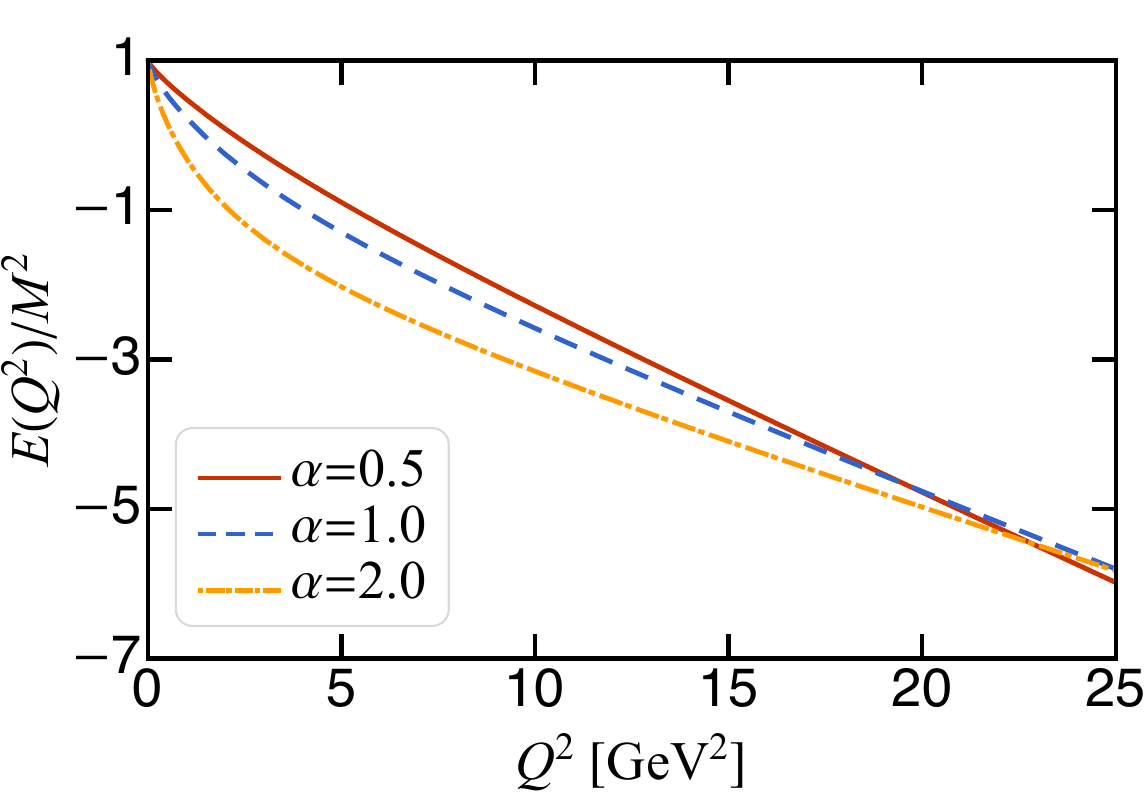}
\caption{Fourier transform of the energy density $e(Q^2)$ and the mass squared density $E(Q^2)$ as a function of $Q^2$. }
\label{fig:energy}
\end{figure}

The pressure distribution is, 
\begin{equation}
p(r_\perp) = \frac{1}{3M} \big(M^2-\mathsmaller{\frac{1}{4}}\nabla^2_\perp\big) \mathcal A(r_\perp) - \frac{1}{3M} \mathcal E(r_\perp)
\end{equation}
And the proper energy density is, 
\begin{equation}
e(r_\perp) = \frac{1}{2M} \mathcal E(r_\perp) + \frac{1}{2M} \big(M^2-\mathsmaller{\frac{1}{4}}\nabla^2_\perp\big) \mathcal A(r_\perp)
\end{equation}
Note that we have introduced several energy related densities~\cite{Lorce:2018egm}, e.g. $e(r_\perp)$, $\mathcal E(r_\perp)$ and $\mathcal T^{+-}(r_\perp)$. 
$e$ is the energy density of the hadron measured in its local rest frame. It is normalized to the hadron rest energy, i.e. hadron mass $M$.  $\mathcal T^{+-}(r_\perp)$ is the light-front energy density multiplied by $2P^+$, the state normalization factor. It is normalized to the hadron light-front energy. $\mathcal E(r_\perp)$ is the one-body invariant mass squared density. It is the light-front energy density in the Breit frame. It is normalized to the hadron invariant mass squared:
\begin{align}
& \int \dd^2 r_\perp \, e(r_\perp) = M, \\
 \frac{1}{2P^+}&\int \dd^2 r_\perp \, \mathcal T^{+-}(r_\perp) = \frac{M^2+P_\perp^2}{P^+}, \\
& \int \dd^2 r_\perp \, \mathcal E(r_\perp) = M^2.
\end{align}
Figure~\ref{fig:energy} shows the energy form factors, i.e., the Fourier transform of the energy densities.

\begin{figure}
\centering
\includegraphics[width=0.45\textwidth]{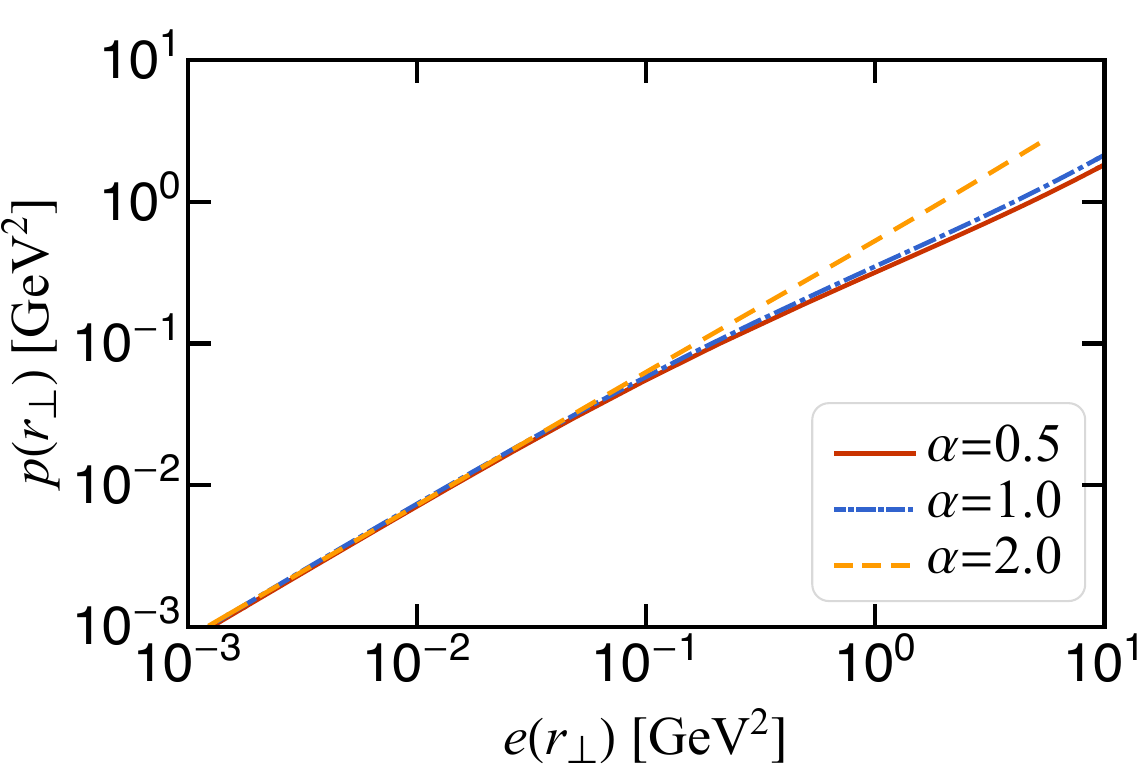}\quad
\includegraphics[width=0.44\textwidth]{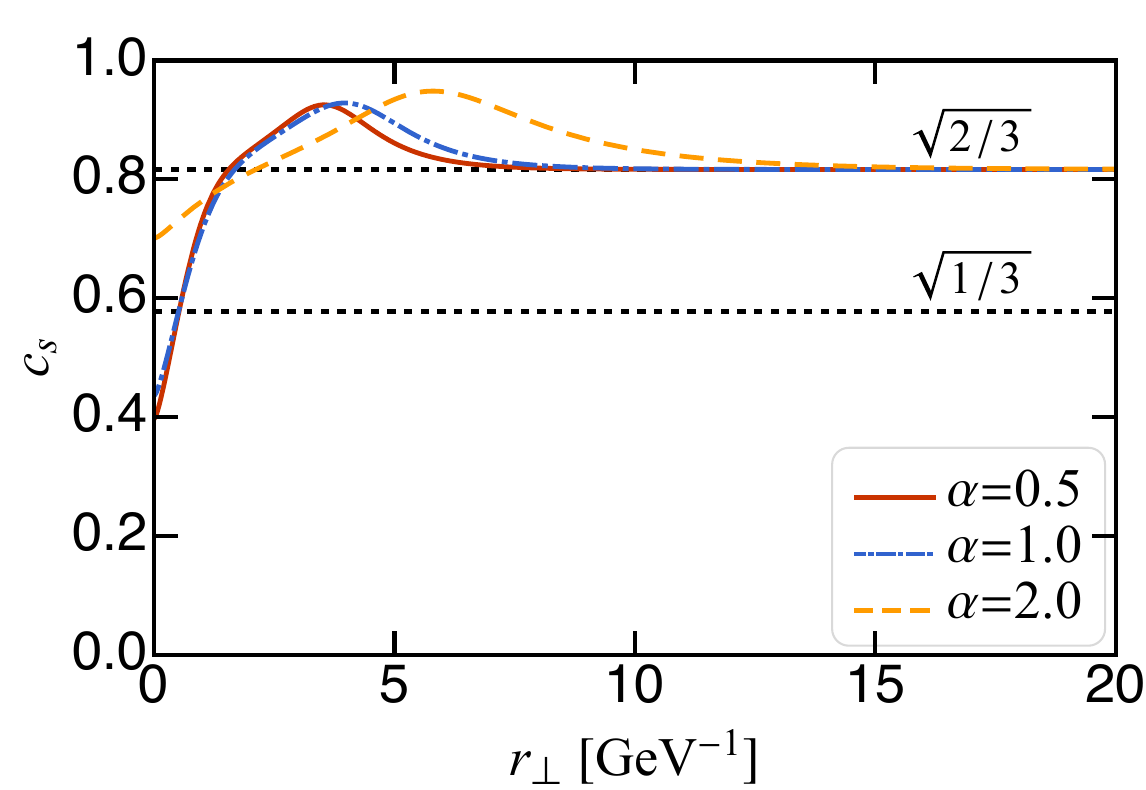}
\caption{(\textit{Left}): The pressure as a function of the energy, i.e., the equation of state for selected couplings. 
(\text{Right}): The distribution of the sound speed $c_s^2 = \dd p / \dd e$ for selected couplings.  }
\label{fig:sound}
\end{figure}

 An interesting quantity to investigate is the the pressure as a function of the energy, which can be interpreted as the equation of state~\cite{Rajan:2018zzy}, as shown in Fig.~\ref{fig:sound}. The derivative of the pressure-energy curve, i.e. $c_s^2 = \dd p / \dd e = p'(r_\perp)/e'(r_\perp)$, is the local speed of the sound. In our model, the speed of the sound exceeds the conformal limit $\sqrt{1/3}$ in a large region~\cite{Altiparmak:2022bke}. At the periphery, it approaches $\sqrt{2/3}$. 

\section{Summary and outlooks}\label{sec:conclusion}

In this work, we computed the forces inside a dressed scalar nucleon in the non-perturbative regime using the light-front Hamiltonian formalism. The calculation is based on a previous non-perturbative solution of the quenched scalar Yukawa model with a systematic Fock sector truncation up to four particles where the Fock sector convergence was demonstrated. The non-perturbative renormalization is implemented using the Fock sector dependent renormalization. In this work, the same counterterms are used to  renormalize the hadronic energy-momentum tensor (EMT). The hadron matrix elements of the EMT are computed up to three particles. 

Instead of using empirical current components, e.g. $T^{11}+T^{22}$, to extract the notoriously challenging $D$ term, we performed a detailed analysis of the Lorentz structure of the EMT and concluded that $T^{+-}$ is consistent with the Hamiltonian dynamics that generates the LFWF of the system, and is a reliable current component for extracting the forces inside the composite particles.  The extracted $D$ term satisfies the von Laue mechanical equilibrium condition. In the forward limit, its value is negative, consistent with the mechanical stability conjecture.$D$ term

Using the higher Fock sector expressions, we not only derived the well-known LFWF representation for the $A$ term, but also obtained a general LFWF expression for the $D$ term.  This expression does not involve the details of the interaction, and can be used to investigate the general properties of the forces inside composite particles in the non-perturbative regime. The expression only involves the diagonal Fock sector contributions and thus can be adapted in phenomenological models to investigate the dynamical structures of the nucleons. For example, for effective interactions between the quark and the antiquark, a reasonable approximation is to couple the graviton to the transverse center of mass of the system $\vec R_\perp = \sum_i x_i \vec r_{i\perp}$, i.e., 
\begin{equation}
\mathcal V_\text{eff}(r_\perp) = \frac{1}{2\pi}\int \frac{\dd x'}{2x'(1-x')} \int \dd^2 r'_\perp \, \widetilde\psi^*(x', \vec r'_{\perp}) 
\delta^2(r_\perp - R_{\perp}) V(r'_\perp) \widetilde\psi(x', \vec r'_{\perp}).
\end{equation}
where, $V(r_\perp)$ is the two-body effective interaction. Note that $\vec R_\perp = 0$ for boost invariant interactions.

An immediate extension of the present work is to incorporate the spin degree of freedom. Another improvement is to lift the quench truncation, which was used to stabilize the scalar theory. Both improvements are required for obtaining a general LFWF representation of the forces in QCD.    

The method we used here can also be applied to investigate the inelastic gravitational form factors, which provide a range of applications from particle production near compact stars, to detecting dark matter, and to gravitational transitions~\cite{Ozdem:2019pkg, Kim:2022bwn, Ozdem:2022zig}. 

\section*{Acknowledgements}

The authors acknowledge fruitful discussions with S. Xu, C. Mondal, S. Nair, Y. Duan, A. Freese, X. Zhao, K. Serafin and Q. Wang. Y.L. thanks V.A. Karmanov for enlightening discussions on the covariant decomposition of the energy-momentum tensor.
Y.L. is supported by the New faculty start-up fund of the University of Science and Technology of China. 
 This work was supported in part by the National Natural Science Foundation of China (NSFC) under Grant No.~12375081, and by the US Department of Energy (DOE) under Grant No.~DE-SC0023692. 

\appendix

\section{Transverse Coordinate Space Representation}\label{sec:coordinate_reprensentation}

Let us introduce the transverse coordinate space wave function as the Fourier transform of the LFWFs, 
\begin{equation}\label{eqn:coordinate_lfwf}
\widetilde \Psi_n(\{x_i, \vec r_{i\perp}\}) = \prod_{i=1}^n \int \frac{\dd^2p_{i\perp}}{(2\pi)^2} e^{-i\vec p_{i\perp}\cdot \vec r_{i\perp}} \Psi_n(\{x_i, \vec p_{i\perp}\}).
\end{equation}
where, $p_{i\perp}$ is the single-particle momentum, and $\Psi$ ($\widetilde\Psi$) is the single-particle momentum-space (coordinate-space) 
light-front wave function. Since the light-front wave functions are boost invariant, the momentum-space wave function can be written in an explicitly boost invariant form, 
\begin{equation}
\Psi(\{x_i, \vec p_{i\perp}\}) = \psi(\{x_i, \vec k_{i\perp}\})
\end{equation}
where, $\vec k_{i\perp} = \vec p_{i\perp} + x_i \vec p_\perp$ is the relative transverse momentum, and $\vec p_\perp = \sum_i \vec p_{i\perp}$ is the total transverse momentum. The $n$-body integration measure can also factorize, 
\begin{equation}
\prod_{i=1}^{n} \int \frac{\dd^2p_{i\perp}}{(2\pi)^2} = \int \frac{\dd^2p_\perp}{(2\pi)^2} \int\big[\dd^2 k_{i\perp}\big]_n
\end{equation}
Here, we have denoted, 
\begin{equation}
\int\big[\dd^2 k_{i\perp}\big]_n = \prod_{i=1}^{n} \int \frac{\dd^2k_{i\perp}}{(2\pi)^2} (2\pi)^2\delta^2(\sum_i k_{i\perp})
\end{equation}
Taking advantage of the light-front boost invariance, Eq.~(\ref{eqn:coordinate_lfwf}) becomes, 
\begin{align}
\widetilde \Psi_n(\{x_i, \vec r_{i\perp}\}) =\,& \int \frac{\dd^2p_{\perp}}{(2\pi)^2}\int \big[\dd^2k_{i\perp}\big] e^{-i\sum_i\vec k_{i\perp}\cdot \vec r_{i\perp}-i\sum_ix_i\vec r_{i\perp}\cdot \vec p_\perp} \psi_n(\{x_i, \vec k_{i\perp}\}) \\
=\,& \delta^2(R_\perp) \int \big[\dd^2k_{i\perp}\big] e^{-i\sum_i\vec k_{i\perp}\cdot \vec r_{i\perp}}\psi_n(\{x_i, \vec k_{i\perp}\}) \\
=\,& \delta^2(R_\perp)\widetilde\psi_n(\{x_i, \vec r_{i\perp}\}) 
\end{align}
Here, we have introduced the intrinsic coordinate-space wave function, 
\begin{equation}\label{eqn:intrinsic_coordinate_lfwf}
\widetilde\psi_n(\{x_i, \vec r_{i\perp}\}) = \int \big[\dd^2k_{i\perp}\big] e^{-i\sum_i\vec k_{i\perp}\cdot \vec r_{i\perp}}\psi_n(\{x_i, \vec k_{i\perp}\}) 
\end{equation}

From Eq.~(\ref{eqn:coordinate_lfwf}), the momentum-space wave function can be expressed as its Fourier transform, 
\begin{align}\label{eqn:momentum_lfwf}
\Psi_n(\{x_i, \vec p_{i\perp}\}) =\,& \prod_{i=1}^n \int \dd^2r_{i\perp} e^{+i\vec p_{i\perp}\cdot \vec r_{i\perp}} \widetilde\Psi_n(\{x_i, \vec r_{i\perp}\}) \\
=\,& \int \big[\dd^2r_{i\perp}\big]_n e^{i\sum_i \vec k_{i\perp} \cdot \vec r_{i\perp}}\widetilde\psi_n(\{x_i, \vec r_{i\perp}\}) = \psi_n(\{x_i, \vec k_{i\perp}\})
\end{align}
Here, again, $\vec k_{i\perp} = \vec p_{i\perp} - x_i \vec p_\perp$ and $\vec p_\perp = \sum_i \vec p_{i\perp}$ is the total momentum. 

Let us next consider the derivatives. From Eqs.~(\ref{eqn:coordinate_lfwf} \& \ref{eqn:momentum_lfwf}), it is not hard to conclude, 
\begin{align}
\vec p_{j\perp} \Psi_n(\{x_i, \vec p_{i\perp}\}) =\,& \prod_{i=1}^n \int \dd^2r_{i\perp} e^{+i\vec p_{i\perp}\cdot \vec r_{i\perp}} i \nabla_{j\perp}  \widetilde\Psi_n(\{x_i, \vec r_{i\perp}\}) \\
i \nabla_{j\perp}  \widetilde\Psi_n(\{x_i, \vec r_{i\perp}\})  =\,& \prod_{i=1}^n \int \frac{\dd^2p_{i\perp}}{(2\pi)^2} e^{-i\vec p_{i\perp}\cdot \vec r_{i\perp}} \vec p_{j\perp} \Psi_n(\{x_i, \vec p_{i\perp}\}).
\end{align}

Since we work with intrinsic variables, let us reduce the both sides of the second expression, 

\begin{align}
\textsc{lhs} =\,& i \nabla_{j\perp}  \widetilde\Psi_n(\{x_i, \vec r_{i\perp}\}) = i \nabla_{j\perp} \delta^2(R_\perp) \widetilde\psi_n(\{x_i, \vec r_{i\perp}\})
+ \delta^2(R_\perp)  i \nabla_{j\perp}  \widetilde\psi_n(\{x_i, \vec r_{i\perp}\}). \\
\textsc{rhs} =\,&  \int \frac{\dd^2p_{\perp}}{(2\pi)^2} e^{-i\vec p_{\perp}\cdot \vec R_{\perp}} \int \big[\dd^2 k_{i\perp}\big]_n e^{-i\vec k_{i\perp}\cdot \vec r_{i\perp}} \big(\vec k_{j\perp} + x_j \vec p_\perp\big) \psi_n(\{x_i, \vec k_{i\perp}\}) \\
=\,&  \int \frac{\dd^2p_{\perp}}{(2\pi)^2} e^{-i\vec p_{\perp}\cdot \vec R_{\perp}} x_j \vec p_\perp  \int \big[\dd^2 k_{i\perp}\big]_n e^{-
i\sum_i\vec k_{i\perp}\cdot \vec r_{i\perp}}  \psi_n(\{x_i, \vec k_{i\perp}\}) \\
\,& +  \int \frac{\dd^2p_{\perp}}{(2\pi)^2} e^{-i\vec p_{\perp}\cdot \vec R_{\perp}}  \int \big[\dd^2 k_{i\perp}\big]_n e^{-i\sum_i \vec k_{i\perp}\cdot \vec r_{i\perp}} \vec k_{j\perp} \psi_n(\{x_i, \vec k_{i\perp}\}) \\
=\,& i\nabla_{j\perp} \delta^2(R_\perp) \widetilde\psi_n(\{x_i, \vec r_{i\perp}\}) 
+ \delta^2(R_\perp) \int \big[\dd^2 k_{i\perp}\big]_n e^{-i\sum_i \vec k_{i\perp}\cdot \vec r_{i\perp}} \vec k_{j\perp} \psi_n(\{x_i, \vec k_{i\perp}\}) 
\end{align}
We can conclude,
\begin{equation}
 i \nabla_{j\perp}  \widetilde\psi_n(\{x_i, \vec r_{i\perp}\}) = \int \big[\dd^2 k_{i\perp}\big]_n e^{-i\sum_i \vec k_{i\perp}\cdot \vec r_{i\perp}} \vec k_{j\perp} \psi_n(\{x_i, \vec k_{i\perp}\})
 \end{equation}
Similarly, let us consider 
\begin{align}
& \int\big[\dd^2r_{i\perp}\big]_n e^{i\sum_i \vec k_{i\perp}\cdot\vec r_{i\perp}} i \nabla_{j\perp}  \widetilde\psi_n(\{x_i, \vec r_{i\perp}\}) \\
=\,& \prod_{i=1}^n \int \dd^2r_{i\perp} e^{i\vec k_{i\perp}\cdot \vec r_{i\perp}} \delta^2(R_\perp) i \nabla_{j\perp}\widetilde\psi_n(\{x_i, \vec r_{i\perp}\}) \\
=\,&  \prod_{i=1}^n \int \dd^2r_{i\perp} e^{i\vec k_{i\perp}\cdot \vec r_{i\perp}} \delta^2(R_\perp) \int \big[\dd^2 k'_{i\perp}\big]_n e^{-i\sum_i \vec k'_{i\perp}\cdot \vec r_{i\perp}} \vec k'_{j\perp} \psi_n(\{x_i, \vec k'_{i\perp}\}) \\
=\,&  \prod_{i=1}^n \int \dd^2r_{i\perp} e^{i\vec k_{i\perp}\cdot \vec r_{i\perp}} \int \frac{\dd^2p_\perp}{(2\pi)^2} e^{-i\vec p_\perp \cdot\vec R_\perp}\prod_{l=1}^n\int \frac{\dd^2 k'_{l\perp}}{(2\pi)^2} (2\pi)^2\delta^2(\sum_l \vec k'_{l\perp}) e^{-i \vec k'_{l\perp}\cdot \vec r_{l\perp}} \vec k'_{j\perp} \psi_n(\{x_l, \vec k'_{l\perp}\}) \\
=\,&  \prod_{i=1}^n \int \dd^2r_{i\perp} e^{i\vec k_{i\perp}\cdot \vec r_{i\perp}} \int \frac{\dd^2p_\perp}{(2\pi)^2} 
\prod_{l=1}^n\int \frac{\dd^2 p'_{l\perp}}{(2\pi)^2} (2\pi)^2\delta^2(\sum_l \vec p'_{l\perp} - \vec p_\perp) e^{-i \vec p'_{l\perp}\cdot \vec r_{l\perp}} (\vec p'_{j\perp}-x_j\vec p_\perp) \Psi_n(\{x_l, \vec p'_{l\perp}\}) \\
=\,&  \prod_{i=1}^n \int \dd^2r_{i\perp} e^{i\vec k_{i\perp}\cdot \vec r_{i\perp}} 
\prod_{l=1}^n\int \frac{\dd^2 p'_{l\perp}}{(2\pi)^2}  e^{-i \vec p'_{l\perp}\cdot \vec r_{l\perp}} (\vec p'_{j\perp}-x_j\underline{\vec p}_\perp) \Psi_n(\{x_l, \vec p'_{l\perp}\}) \\
=\,&  (\vec k_{j\perp}-x_j\underline{\vec p}_\perp) \psi_n(\{x_i, \vec k_{i\perp}\}) \\
=\,&  \vec k_{j\perp}  \psi_n(\{x_i, \vec k_{i\perp}\}) 
\end{align}
where, $\underline{\vec p}_\perp = \sum_l \vec p'_{l\perp}$. In the last equality, we have used the fact that $\sum_i \vec k_{i\perp} = 0$. We have used a change of variable: $\vec k'_{l\perp} = \vec p'_{l\perp} - x_l \vec p_\perp$. 
Therefore, we can conclude, 
\begin{equation}
\vec k_{j\perp} \psi_n(\{x_i, \vec k_{i\perp}\}) = \int \big[\dd^2 r_{i\perp} \big]_n e^{+i\sum_i\vec k_{i\perp}\cdot \vec r_{i\perp}} i \nabla_{j\perp}\widetilde\psi_n(\{x_i, \vec r_{i\perp}\})
\end{equation}

We write the $n$-body integration measure as, 
\begin{equation}
\int\big[ \dd x_i \dd^2k_{i\perp} \big]_n = \frac{1}{S_n}\prod_{i=1}^{n}
\int \frac{\dd x_i}{2x_i}\frac{\dd^2k_{i\perp}}{(2\pi)^3} 
2\delta(\sum_i x_i - 1) (2\pi)^3\delta^3(\sum_i \vec k_{i\perp} ) \equiv \frac{1}{S_n}\int \big[\dd x_i \big]_n \big[\dd^2k_{i\perp}\big]_n
\end{equation}
where, $S_n$ is the symmetry factor. The longitudinal measure is, 
\begin{equation}
\int \big[\dd x_i \big]_n \equiv \prod_{i=1}^{n}
\int \frac{\dd x_i}{4\pi x_i} 
4\pi \delta(\sum_i x_i - 1) 
\end{equation}
Recall, the transverse measure is, 
\begin{equation}
\int \big[\dd^2 k_{i\perp} \big]_n \equiv \prod_{i=1}^{n}
\int \frac{\dd^2 k_{i\perp}}{(2\pi)^2} 
2\pi \delta^2(\sum_i \vec k_{i\perp} ) 
\end{equation}
We also define a new $n$-body measure based on the transverse coordinates, 
\begin{equation}
\int\big[\dd x_i \dd^2 r_{i\perp} \big]_n \equiv \frac{1}{S_n}\int \big[\dd x_i\big]_n \big[\dd^2r_{i\perp}\big]_n 
= \frac{1}{S_n} \prod_{i=1}^n \int \frac{\dd x_i}{4\pi x_i} \dd^2r_{i\perp} 4\pi \delta(\sum_i x_i - 1) \delta^2(\sum_i x_i \vec r_{i\perp})
\end{equation}


\end{document}